\documentclass[twocolumn, prb]{revtex4}

\usepackage{graphicx}
\usepackage{amssymb}
\usepackage{epstopdf}
\usepackage{amsmath}
\usepackage{color}
\makeatletter

\@ifundefined{textcolor}{}
{
 \definecolor{BLACK}{gray}{0}
 \definecolor{WHITE}{gray}{1}
 \definecolor{RED}{rgb}{1,0,0}
 \definecolor{GREEN}{rgb}{0,1,0}
 \definecolor{BLUE}{rgb}{0,0,1}
 \definecolor{CYAN}{cmyk}{1,0,0,0}
 \definecolor{MAGENTA}{cmyk}{0,1,0,0}
 \definecolor{YELLOW}{cmyk}{0,0,1,0}
}

\makeatother

\begin{document}

\title{Majorana Spin Liquids, Topology and Superconductivity in Ladders}

\author{Karyn Le Hur$^{1}$\footnote{The authors are placed by alphabetical order.}, Ariane Soret$^{1,2}$, Fan Yang$^{1}$}
\address{$^{1}$ Centre de Physique Th\' eorique, \'Ecole Polytechnique, CNRS, Universit\' e Paris-Saclay, 91128 Palaiseau Cedex, France \\
$^{2}$ Department of Physics, Technion Israel Institute of Technology, 32000 Haifa, Israel}

\date{\today}

\begin{abstract}
We theoretically address spin chain analogs of the Kitaev quantum spin model on the honeycomb lattice. The emergent quantum spin liquid phases or  Anderson resonating valence bond (RVB) states can be understood, as an effective model, in terms of p-wave superconductivity and Majorana fermions. We derive a generalized phase diagram for the two-leg ladder system with tunable interaction strengths between chains allowing us to vary the shape of the lattice (from square to honeycomb ribbon or brickwall ladder). We evaluate the winding number associated with possible emergent (topological) gapless modes at the edges. In the $A_z$ phase, as a result of the emergent $\mathbf{Z}_2$ gauge fields and $\pi$-flux ground state, one may build spin-1/2 (loop) qubit operators by analogy to the toric code. In addition, we show how the intermediate gapless $B$ phase evolves in the generalized ladder model. For the brickwall ladder, the $B$ phase is reduced to one line, which is analyzed through perturbation theory in a rung tensor product states representation and bosonization.  Finally, we show that doping with a few holes can result in the formation of hole pairs and leads to a mapping with the Su-Schrieffer-Heeger model in polyacetylene;  a  superconducting-insulating quantum phase transition for these hole pairs is accessible, as well as related topological properties. 
\end{abstract}

\maketitle

\section{Introduction}

The quest for topological phases has attracted some attention recently in relation with exotic quantum states of matter related to Chern insulators \cite{Duncan} as well as topological insulators and superconductors \cite{Konig,HasanKane,Stanford,Bernevig}. The energy spectrum is characterized by non-trivial Bloch bands and by a topological index \cite{TKNN}. At the same time, this index is related to the occurrence of protected chiral edge modes, by analogy with the quantum Hall effect \cite{Hall,Laughlin,Bertrand}, due to a bulk-edge correspondence. It is also important to mention related progress in ultra-cold atoms and photon systems where one can artificially engineer similar quantum phases \cite{Dalibard,Soljacic,CRAS}. In addition, the topological invariants, such as Chern number or Zak phase can be measured with very high accuracy \cite{Munich1,Munich2,Hamburg,Ian}. Berry phases \cite{Berry} have also been measured in high-$T_c$ superconductors and graphene \cite{Gervais,Orenstein,Kim} and in superconducting quantum circuits \cite{PolkovnikovGritsev,Review,Lehnert,Roushan,AndreasBerry,CRAS}. Quantum materials are also characterized by intrinsic interactions, and therefore extending the notion of topological phases to interacting band structures seems a timely subject of interest. Theoretical progress \cite{PesinBalents,Stephan,TianhanBenoit,Lee} based on slave-rotor, quantum field theory techniques \cite{FlorensGeorges} and numerical approaches \cite{WeiDMFT,Wurzburg,Walter} have been accomplished and are also related to the discovery of quantum materials \cite{Kee} and to the engineering of Feynman quantum simulators \cite{Feynman,Ignacio,Immanuel,AHJ,CRAS2017}. Increasing interactions also naturally connects these states of matter to Mott physics and possible quantum spin liquids which do not exhibit long-range order and are related to Anderson Resonating Valence Bond (RVB) States \cite{AndersonRVB,AndersonScience,LDA}.  An important example of quantum spin liquid ground state emerges in the Kitaev model on the honeycomb lattice \cite{Kitaev06}, which can be solved exactly and bridges between the occurrence of Majorana particles in the ground state and the possibility to realize protected quantum information operations \cite{Nayak,Mong} through braiding these Majorana particles. Generalizations in three dimensions have also been addressed \cite{Maria,Oxford} as well as in models with long-range (and disordered) forces \cite{SachdevYe,Qi}. We note the discovery of recent quantum materials related to the Kitaev model \cite{Jackeli,Kee,Dima,Simon}. This research is also linked to the search of quantum spin liquid states and superconductivity on Kagome materials \cite{Mendels,Cava,Roser}. Majorana fields and particles have also been predicted in high-energy physics (in the context of neutrinos), nuclear physics \cite{Franz}, and recently in relation with the Sachdev-Ye-Kitaev model \cite{SachdevYe,Kitaevnew,Polchinsky,Witten}. 

In this paper, we address chain networks of Kitaev quantum spin models with a $\mathbf{Z}_2$ symmetry \cite{Feng,Wu,Smitha,Motrunich,Loss,Yao} (see Fig. 1). Through the Jordan-Wigner transformation \cite{Feng}, the Kitaev spin chain is related to an effective Bardeen-Cooper-Schrieffer (BCS) model for superconductivity.  From a theoretical point of view, quasi-one-dimensional systems with Heisenberg coupling also offer analytical solutions to connect quantum spin liquids and superconducting ground states  \cite{LBF,Brookhaven} with potential relevance to the physics of the cuprates \cite{ZhangRice,KarynMaurice}.  These wire constructions are therefore important to study the link between quantum spin liquid or spin system with short-range magnetic interactions and the occurrence of superconductivity \cite{Kivelson,RK,MSF}. 

More precisely, the Kitaev magnetic chain \cite{Feng} yields an emergent BCS Hamiltonian with a p-wave pairing symmetry, making an analogy with the physics of Helium-3 \cite{AndersonMorel,AndersonBrinckmann,Balian,Leggett,Volovik}, topological superconductors \cite{Read,Boulder} and topological superconducting quantum wires \cite{Kitaev,Oreg,Sarma,Leo,Alicea,Bert,Matthew}. Similar topological ferromagnetic chains have already been studied and engineered by proximity effect with a superconductor with spin-orbit coupling \cite{Yazdani,Felix}. In two dimensions, emergent superconductivity has also been predicted theoretically in (doped) magnetic Kitaev models \cite{Burnell,Rosenow,Tianhan}. A Kitaev magnetic chain can also be seen as a strong-coupling analogue of the Su-Schrieffer-Heeger (SSH) model in polyacetylene \cite{SSH,Asboth}, which possesses two different tunneling parameters in the Hamiltonian and can show the emergence of topological edge excitations, by analogy to the Fibonacci chain \cite{LPNMarcoussis}. Localized topological soliton states have been recently observed in ultra-cold atoms \cite{Bryce}. In the spin chain language, this will traduce the possible emergence of gapless spin excitations at the edges by analogy with the spin-1 chain \cite{Haldanespin1,AKLT,spin1TO,QInfo}. Related Hamiltonians, such as the Rice-Mele model, have also been engineered in ultra-cold atoms and photon systems and topological properties have been measured \cite{Munich2,Munich,Kitagawa}. The related Jackiw-Rebbi model \cite{Jackiw} can also be realized \cite{Angelakis}. 

Before addressing the main objectives and results of the paper, it is perhaps important to mention possible realizations of such  Kitaev spin Hamiltonian \cite{Kitaev06,Feng}. In principle, these spin chains can be simulated based on existing proposals and technology in ultra-cold atoms \cite{Duan}, Circuit Quantum Electrodynamics and Josephson circuit architectures \cite{SPEC}, where the coupling between qubits (for example, transmon qubits) can be of $X$ or $Y$ type in principle \cite{Martinischain,Martinis,Schmidt}.  In circuit quantum electrodynamics architectures, the Ising X coupling would correspond to a capacitive coupling and Y to an inductive coupling between transmon qubits. Realizing a spin chain with alternating couplings $XYXYX...$ seems achievable \cite{Martinischain}. An Ising $Z$ coupling between Kitaev chains would correspond to an interaction (Kerr) coupling between chains in the equivalent Bose-Hubbard representation between transmon qubits \cite{Martinischain}. At  a general level, these interaction terms can be simulated by non-linearities (cross-Kerr effect) of the Josephson elements in superconducting chains \cite{Neel,Fluxonium}. Engineering of other quantum spin chains with $\mathbf{Z}_2$ symmetry, such as the quantum Ising chain, using Josephson junctions and superconducting elements was proposed in Ref. \onlinecite{Mooij}, in relation with the emergence of topological qubits. The realization of topological qubits in these superconducting circuits have also started to attract some attention experimentally \cite{Rutgers,Ioan}. Similar progress has been realized in ultra-cold atoms to engineer spin chains \cite{Greiner,Greiner2} and Resonating Valence Bond states \cite{atomRVB}.

\begin{figure}[t]
\centering
\includegraphics[scale=0.6]{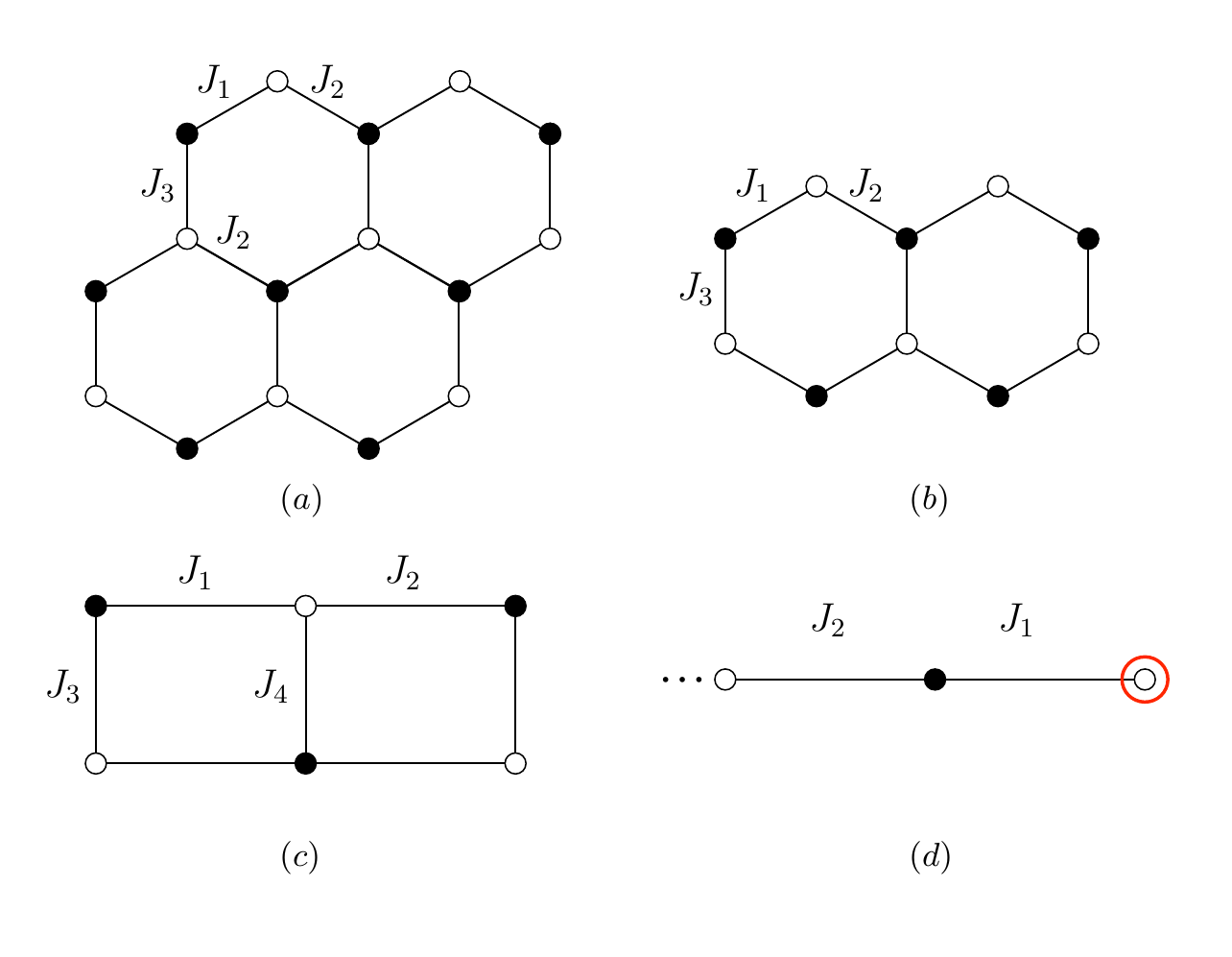}
\includegraphics[width=0.45\textwidth]{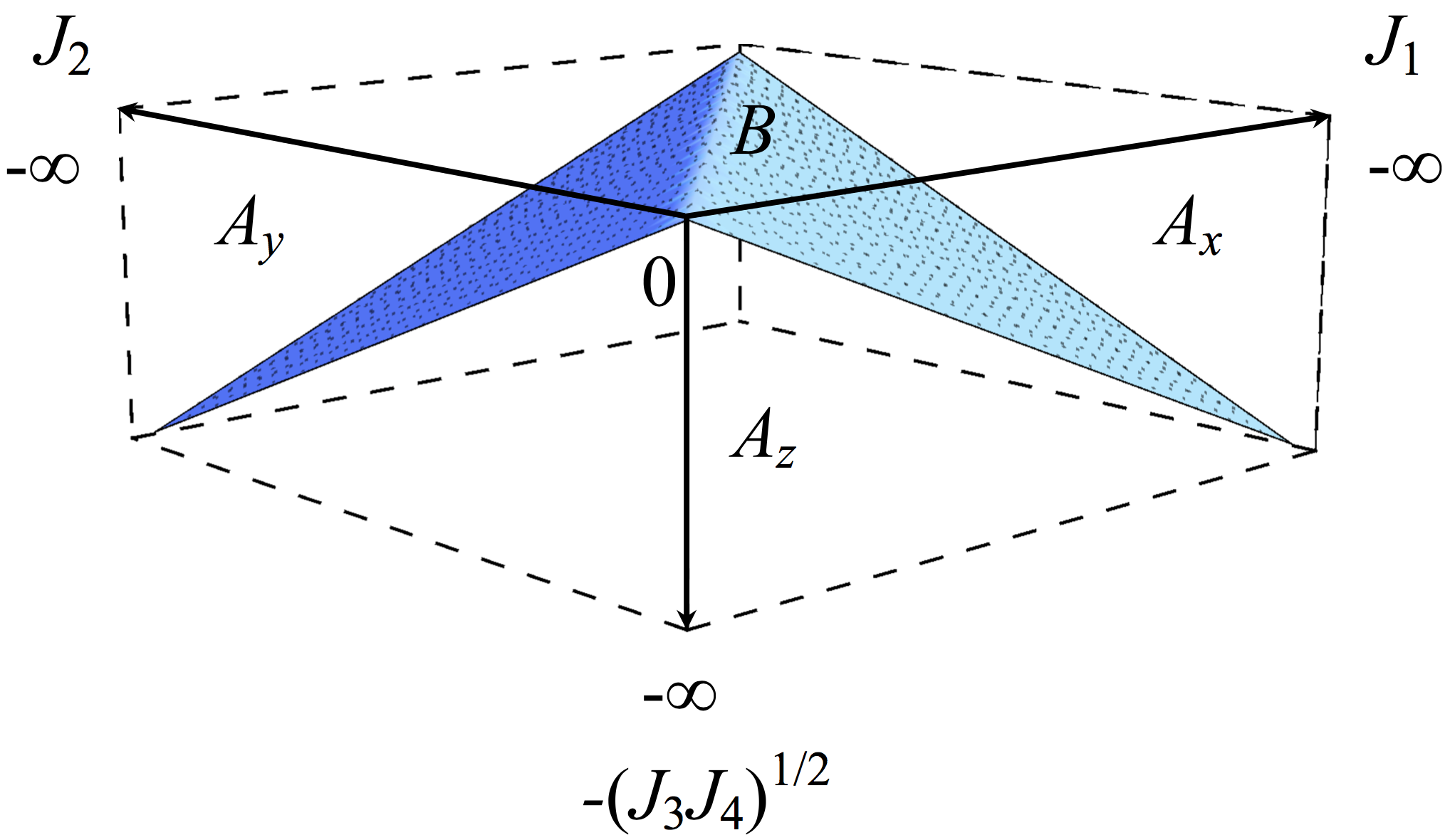}
\caption{Different Geometries: Honeycomb Lattices (a) to Brickwall (b) and Square (c) Ladders, and Chains (d). (In the limit of large system sizes, the figure (b) which constitutes a honeycomb ribbon is similar to a brick wall ladder and the figure (c) which is a rectangular ladder is similar to a square type ladder. Bottom:  Our generalized phase diagram for the quantum ladder (which will be discussed in
detail in Sec. III).}
\label{cartoon}
\end{figure}

Now, we summarize the main objectives and results of the paper. We study different geometries, related to Fig. 1, starting from a single chain (Fig. 1d) and then exploring ladder systems such as the Square ladder (Fig. 1c) and the Brickwall ladder (Fig. 1b) which can be viewed as a Kitaev honeycomb ribbon geometry (Fig. 1a). One goal of the paper is to extend the analysis of Ref. \onlinecite{Feng} (addressing the Square ladder) to a generalized ladder geometry (between the square and the brickwall ladder) by varying the parameters $J_3$ and $J_4$. In particular, to the best of our knowledge, the brickwall ladder has not been addressed previously. Next, we will study  doping effects with a pair of holes and derive an effective SSH model. Using the Majorana representation, we evaluate rigorously the phase diagrams. As a reminiscence of the two-dimensional Kitaev model \cite{Kitaev06}, we identify gapped spin liquid phases with possible short-range spin order on strong nearest-neighbor links  along the $X$ direction in the $A_x$ phase, along the $Y$ direction in the $A_y$ phase, or along the $Z$ direction in the $A_z$ phase. The $A_x$ and $A_y$ phases already occur in single-chain configurations. Emergent gapped spin-liquid phases can be characterized by a topological string order parameter \cite{Feng} by analogy to spin-1 chains \cite{Haldanespin1,AKLT,spin1TO,QInfo}. The possible existence of gapless excitations in the $A_x$ and $A_y$ magnetic phases can also be described through a quantized winding number by analogy to the SSH model. We discuss the nature of such edge modes (spin-1/2 objects, Majorana fermions).

For the Brickwall ladder, following the phase diagram in Fig. 1 with $J_3$ or $J_4=0$, we identify a line of gapless spin excitations separating two gapped phases with different spin polarizations on strong bonds. This line of excitations can be seen as a precursory effect of the $B$ phase in the two-dimensional Kitaev model on the honeycomb lattice \cite{Burnell} and it will spread out in the generalized ladder phase diagram (Fig. 1). In a gapped phase, the spin polarization on these strong bonds can vary continuously from the $X$ to $Z$ axis or from the $Y$ to $Z$ axis. Using bosonization and perturbation theory, along the line of gapless spin excitations, we will show the occurrence of pre-formed pairs in the system. For large inter-chain coupling, the magnetic system can be seen as a matrix product states (or rung tensor product states) representation \cite{Cirac}. Then, we study the propagation of a hole pair along this line of gapless excitations. We show that these hole pairs can be precisely described by a SSH model in the intermediate regime of inter-chain coupling, with the occurrence of two tunneling amplitudes for the brickwall ladder. We identify two possible phases;  a hole pair can localize at the edges of the ladder or the hole pair can coherently propagate along the chains at weaker inter-chain coupling, forming a quasi-one-dimensional superconductor. The localized phase for the hole pair is characterized by a quantized winding number by analogy to the SSH model. For the Square ladder, this analysis with two holes suggests that the system in the dilute hole limit can be seen as a quasi-one-dimensional superconducting spin liquid starting from the $A_z$ phase. 

The organization of the paper is as follows. In Sec. II, we summarize briefly known properties of the single chain system (to fix the notations that will be useful to study next sections), and we discuss gapless edge and bulk excitations. In Sec. III, we study the ladder systems using the Majorana representation and build our phase diagram of Fig. 1. Then, we study in more details the brickwall ladder phase diagram, which has not been studied before. In addition, we study the possibility to build Majorana qubit loop operators in these ladders. In Sec. IV, we study the quantum phase transitions in more detail, in particular for the brick wall ladder, where we identify a gapless line in the phase diagram. First, we apply a perturbation theory along this gapless line in the phase diagram for the Brickwall ladder and show the occurrence of pre-formed pairs in the system by using a rung tensor product representation. Then, we apply a bosonization approach to reinforce the notion of pre-formed pairs in the system along the line of gapless excitations. We also predict that similar gapless excitations exist in the Square ladder, at the transition lines between the $A_z$ phase and the $A_x$ and $A_y$ phases, respectively. In Sec. V, we study the propagation of a pair of holes in the Brickwall ladder system and address the possible emergence of a topological insulator - superconducting transition for a hole pair. In Sec. VI, we present a summary of the results and discuss relevance for current experiments. In Appendix A, we give some details on Fourier transform and winding number calculations related to the magnetic phases. In Appendix B, we show how braiding of two nearest neighbor Majorana fermions could be implemented in relation with Sec. IIID. In Appendix C, we present a derivation of the Hamiltonian for the ladder system using different string configurations, to show the ``gauge-invariant form'' (or string-invariant form) of the Hamiltonian. In Appendix D, we present a spin-spin correlation function analysis related to our ladder phase diagram in Fig. 6. In Appendix E, we present the renormalization group analysis related to Sec. IVB.

\section{Kitaev Spin Chain}

In Sec. IIA, we briefly summarize known properties of the one-dimensional (1D) magnetic Kitaev chain. In Sec. IIB, we analyze the presence of edge excitations in spin and Majorana
representations. The calculation of the winding number in the the Anderson pseudo-spin representation \cite{Anderson} is presented in Appendix A, making an analogy with the
SSH model in the Mott regime. We also relate Majorana excitations in the spin chain with the degeneracy of the ground state. In particular, the Kitaev spin chain also reveals a chain of gapless Majorana excitations, which can be used in network devices (as illustrated in Sec. III) to encode information in ${\cal Z}_2$ variables \cite{Terhal,Fu,AltlandEgger,Flensberg}. We also suggest possible implementations of  Majorana braiding, by analogy to topological superconducting wires \cite{Kitaev,Oreg,Sarma,Leo,Alicea,Bert}.

In this model, assuming we start with a Mott insulating phase, spin-$\frac{1}{2}$ particles are located on the vertices (sites) of a chain and interact with nearest neighbors. The interactions are supposed to be ferromagnetic $J_1,J_2\leq 0$ (however, we will notice an invariance of the energy spectrum under the transformation $J_i \rightarrow -J_i$, simultaneously for $i=1$ and $i=2$), and the links or magnetic couplings are of two types : "$x$-links" (Ising interaction along the $X$ direction) and "$y$-links" (Ising interaction along the $Y$ direction) alternatively (see Fig. 1). The corresponding Hamiltonian is \cite{Feng,Kitaev06}:

\begin{equation}
H=\sum_{j=2m-1}J_1\sigma_j^x\sigma_{j+1}^x+J_2\sigma_{j+1}^y\sigma_{j+2}^y.
\end{equation}
With this notation, the sum runs over odd sites only, such that $m\geq 1$ is an integer. To compute the winding number associated with the spin chain in Appendix A, by analogy with the SSH model, we will use a spin chain  which possesses $2M$
sites and $M$ unit cells. In that case, the spin chain also finishes with a $J_1$ link (see Fig. (1d)).

\subsection{Properties and Known Results}

To make a connection with the BCS model, one can rewrite the Hamiltonian in the fermionic representation, where the quantum spin operators are replaced by fermionic operators $(a^{\dagger}, a)$ using the Jordan-Wigner transformation \cite{JW}: 
\begin{equation}
\left\{
\begin{array}{ll}
      \sigma_{j}^x = (a_{j}^\dagger+ a_{j})e^{i\pi\sum_{\{i\}\in string}a_{i}^\dagger a_{i}} \\
      \sigma_{j}^y = -i(a_{j}^\dagger- a_{j}) e^{i\pi\sum_{\{i\}\in string}a_{i}^\dagger a_{i}}\\
      \sigma_{j}^z = 2 a_{j}^\dagger a_{j}-1=
\left\{
\begin{array}{ll}
 1 \mbox{, if $|\uparrow_j\rangle _z$}\\
-1 \mbox{, if $|\downarrow_j\rangle _z$}.
\end{array}
\right.							
    \end{array}
\right.
\end{equation}
(For simplicity, eigenvalues of the spin-1/2 are normalized to $+1$ and $-1$; this representation introduced in 1928 maps quantum spin operators obeying the Lie algebra to spinless fermionic operators with occupancy $0$ and $1$ at a site). We choose a certain path to define the string operator, stopping at site $j-1$. For example, for the single chain, we start the string on the left of the chain at site $m=1$. 

In this representation, the Hamiltonian turns into :
\begin{eqnarray}
H = \sum_{j=2m-1}  J_1 (a_{j}^\dagger- a_{j})(a_{j+1}^\dagger+ a_{j+1}) \\ \nonumber
- J_2(a_{j+1}^\dagger+ a_{j+1})(a_{j+2}^\dagger- a_{j+2}).
\end{eqnarray}
In the absence of applied magnetic field,  then $\langle \sigma_j^z\rangle =0$, i.e. $\langle a_{j}^\dagger a_{j}\rangle =\frac{1}{2}$. Equivalently, the effective chemical potential for the spinless fermions in Eq. (3) is $\mu=0$. 

We impose periodic boundary conditions and perform a Fourier transform to access the bulk properties of the system:
\begin{eqnarray}
H = \sum_k  (J_1+J_2) \cos(kl) (a_{k}^\dagger a_{k} - a_{k} a_{k}^{\dagger}) \\ \nonumber
+ i (J_1-J_2)\sin(kl)(a_{-k}^\dagger a_{k}^\dagger+ a_{-k} a_{k}),
\end{eqnarray}
which is equivalent to the BCS Hamiltonian, with a pairing term of the form $\Delta_k \propto i(J_1-J_2)\sin(kl)$. We note the analogy with one-dimensional p-wave superconductors \cite{Kitaev}. Within our definitions of the Fourier transform and wave-vectors, we derive the energy spectrum (in Appendix A, we check this result using the Majorana fermion representation used in Sec. IIB): 
\begin{equation}
\epsilon(k)=\pm \sqrt{J_1^2+J_2^2+2J_1J_2\cos(2kl)}.
\end{equation}
In this formulation the wave-vector is defined in the reduced Brillouin zone as $-\pi/(2l) \leq k \leq \pi/(2l)$ implying that the energy spectrum is invariant under the transformation $2kl \rightarrow 2kl + 2\pi$.  At the quantum phase transition $J_1=J_2$, we note that this convention is in agreement with $\cos(2k l)=-1$ meaning a gap closure at the Fermi wave-vector $k_F=\pi/(2l)$. The Hamiltonian describes a half-filled band for the $a_k$ and $a_k^{\dagger}$ fermions, which is imposed by the chemical potential $\mu=0$. If we fix the lattice spacing $2l$ to unity, then this gives $-\pi \leq k \leq \pi$ in agreement with Refs. \onlinecite{Feng,Motrunich,Loss}. One thus establishes an exact mapping between the 1D magnetic Kitaev chain and the BCS Hamiltonian. From the expression of $\epsilon(k)$ above, it is clear that the spectrum is gapped for $J_1\neq J_2$, and gapless for $J_1=J_2$. 

\begin{figure}[t]
\centering
\includegraphics[scale=0.9]{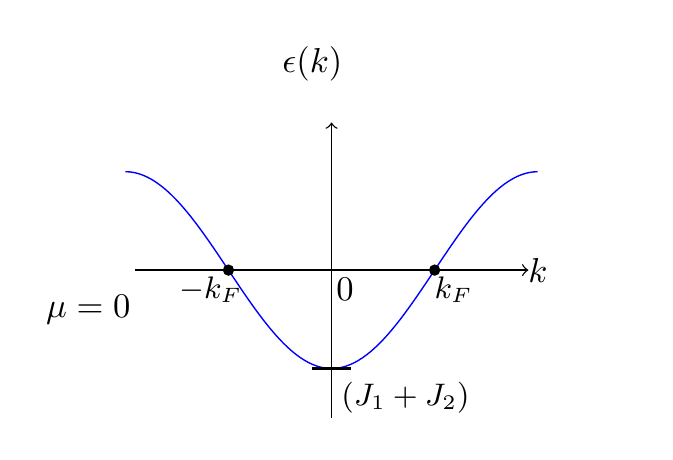}
\includegraphics[scale=0.9]{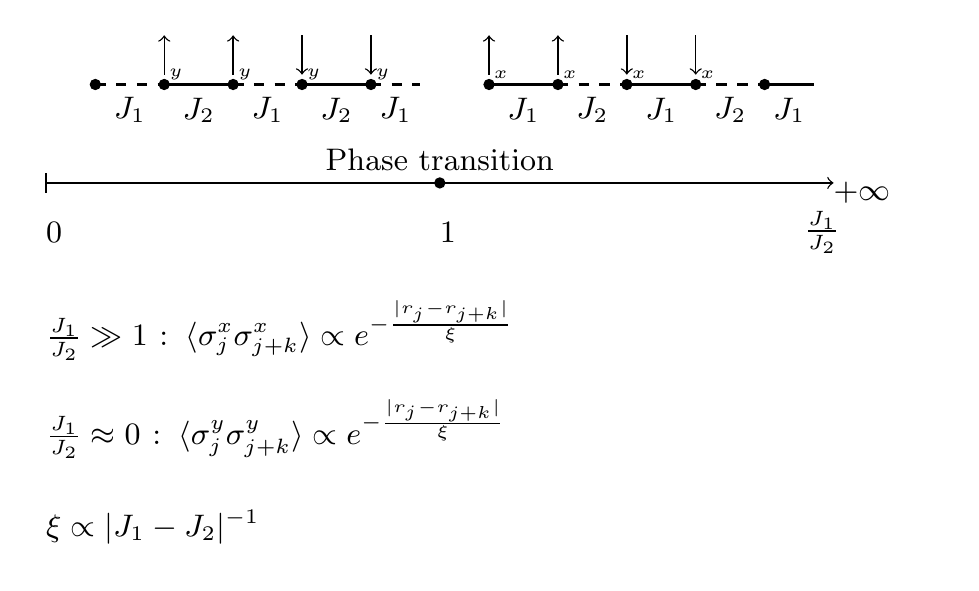}%
\caption{(color online) Simple fermionic dispersion relation for the 1D chain model at $J_1=J_2$. The gapped phases at $\frac{J_1}{J_2}>1$ and $\frac{J_1}{J_2}<1$ are spin liquids, characterized by a diverging coherence length $\xi\propto |J_1-J_2|^{-1}$ close to the phase transition (traducing a power-law behavior of these correlation functions for $J_1=J_2$). In the gapped phase, there is the formation of Valence Bonds between nearest neighbors, which can resonate when approaching the quantum phase transition. This description then shows some analogy with the spin-1 chain construction \cite{AKLT}. Similar resonating valence bond descriptions have been analyzed to connect N\' eel and dimer phases in two dimensions \cite{LDA}. These dimers can also be seen as pre-formed p-wave superconducting pairs as in Helium-3 \cite{AndersonMorel,AndersonBrinckmann,Balian,Leggett}.}
\label{phase_trans_1D}
\end{figure}

The gapped phases associated with a BCS (p-wave like) pairing term between the Jordan-Wigner fermions are in fact RVB spin liquids (corresponding to the $A_x$ and $A_y$ phases described by Kitaev \cite{Kitaev06}), characterized by exponentially decreasing correlation functions. Deep in the $A_x$ or $A_y$ phase the correlation length converges to the lattice spacing. In Fig. 2, nearest neighbor sites $j$ and $j+1$ coupled with a strong link (coupling) can be either $|\uparrow_j \uparrow_{j+1}\rangle=|+_j +_{j+1}\rangle$ or 
$|\downarrow_{j} \downarrow_{j+1}\rangle=|-_j -_{j+1}\rangle$ following the $x$ $(y)$ axis in the $A_x$ $(A_y)$ phase and there is no correlation between these bonds when $J_2\rightarrow 0$  $(J_1\rightarrow 0)$.  The quantum degeneracy in the chain associated with these bonds in the $A_x$ phase (or $A_y$ phase) then is $2^{M}$ in the thermodynamical limit, and $M$ is the number of bonds coupled by $J_1$. The right (and left) boundary of the lattice can produce an extra spin-1/2 excitation (see Fig. 1 and Sec. IIB). Far in a gapped spin liquid phase, for example in the $A_x$ phase, one can check that the $J_2$ coupling cannot induce a long-range Ising order. However, valence bonds can resonate in principle by application of the $J_2$ coupling, which produces virtual excitations described through the cross-term in $J_1 J_2$ in Eq. (5). Other descriptions in terms of non-local string order parameters and an emergent dual quantum Ising model are possible to describe the quantum phase transition and the macroscopic degeneracy (odd or even sites decouple) \cite{Feng}. 

At $J_1=J_2$, a phase transition occurs, and in the fermion representation the ground state corresponds to a  Fermi sea 
\begin{equation}
|GS\rangle = \prod_{k<k_F} a_k^{\dagger}|0\rangle,
\end{equation}
 with $k_F=\pi/(2l)$ (such that $\epsilon(k_F)=\mu=0$) characterized by the band structure of free electrons $\epsilon(k)=2J_1\cos(kl)$ plotted in Fig.~3. Essentially, the pairing terms in $a^{\dagger}_k a^{\dagger}_{-k}$ become zero and the tight-binding Hamiltonian in Eq. (3) is equivalent to that of a tight-binding model with a single hopping amplitude $-J_1$ and a lattice spacing $l$. Since the energy spectrum of the fermions is linear around the chemical potential $\mu=0$ in Fig. 2, this will allow us to apply a bosonization approach \cite{Haldane,Giamarchi} in Sec. IVB when switching on a small coupling between the chains. At the quantum phase transition, spin-spin correlation functions decay as power laws both at $k=0$ and $2k_Fl=\pi$. We also infer based on the recent Ref. \onlinecite{Loic_fluctuations}, that bi-partite spin fluctuations could detect the quantum phase transition and associated gapless excitations at the edges. Similar proposals have been suggested by coupling to a cavity field; see for example Ref. \onlinecite{Olesia}. Such bi-partite fluctuations have been shown to be useful to describe many-body systems and quantum phase transitions \cite{Stephanfluc,Francisfluc,Alex}. We also note recent progress to observe such bi-partite fluctuations based on correlation functions \cite{Greiner2}. The phase diagram and the physical properties of the phases are summarized in Fig.~\ref{phase_trans_1D}.  
 
 This BCS representation will be very useful to study the Brickwall ladder system (and also the Square ladder) at small $J_3$ in Sec. IVB, where the bosonization approach will show an analogy from quantum field theory, between the two-leg Kitaev spin ladder and an emergent quasi-one dimensional superconductor of charges $2e$ \cite{LutherEmery}. This also makes an analogy with superconductivity in quasi-one-dimensional materials \cite{Jerome,KarynMaurice} and Resonating Valence Bond States of Anderson \cite{AndersonRVB}. A quantum field theory description was also developed in two dimensions to connect Kitaev spin liquids and emergent superconducting Hamiltonians \cite{Burnell}.

\subsection{Edge and Bulk gapless excitations}

Here, we study in more detail gapless excitations in the bulk and at the edges. Our objective is to study gapless excitations both in the spin and Majorana representations, complementing the efforts in Refs. \onlinecite{Smitha,Loss}.  

In Appendix A, by analogy to the SSH model \cite{SSH} with $2M$ sites, we evaluate
the winding number associated with the edge excitations using the Anderson pseudo-spin representation \cite{Anderson} of Ref. \onlinecite{Wu}. The SSH model belongs to the topologically protected symmetry class BDI \cite{BernevigHouches} (the presence of edge modes can be inferred from the momentum distribution function \cite{Carlos} and from bipartite fluctuations \cite{Loic_fluctuations,Stephanfluc,Francisfluc,Alex}). Based on the results of Appendix A, the $A_x$ phase has no gapless excitations (and a winding number zero) whereas the $A_y$ phase yields edge excitations at both edges and a winding number $1$ (which can be understood as spin-1/2 edge excitations in the limit $J_1\rightarrow 0$ in Fig. 1). In two dimensions, the gapped phases $A_x$ and $A_y$ do not reveal edge modes (the Chern number is zero) \cite{Kitaev}.  Switching on a perturbative $J_1$ coupling one can check that the edge modes  only couple to order  $\sim (J_1/J_2)^M\sim \exp(M\ln (J_1/J_2))$ where $J_1/J_2\rightarrow 0$. 

More precisely, if we start with a spin at an edge in the state $|+\rangle_y$ and the nearest neighbor bond is in a state $|++\rangle_y$, then the application of the $J_1$ coupling turns the state of the three spins $|+\rangle_y \otimes |++\rangle_y$ into $|-\rangle_y \otimes |-+\rangle_y$, which corresponds to an excited state separated by $2|J_2|$ from the ground state. This argument can be repeated (and generalized to another preparation state) and the only non zero order in the perturbation theory then should couple the two edges, whereas the bulk states return to the ground state. This argument will also apply for the ladder system described below in the $A_x$ and $A_y$ phases, and in Sec. V for the SSH model of the hole pair. In the ladder system studied below, the edge modes will occur on a single chain for the two phases $A_x$ and $A_y$. The exponential suppression of the coupling between edge modes for large $M$ also reflects that the chain can be described by a non-local string order parameter \cite{Feng} by analogy to the spin-1 chain. 

It is now useful to rewrite the Kitaev spin chain in a Majorana fermion language and re-analyze the ground state properties. More precisely,
\begin{eqnarray}
      d_j &=& (a_j^{\dagger}+a_j) \nonumber \\
      c_j &=& i(a_j^{\dagger}-a_j),	
\end{eqnarray}
such that $d_j^{\dagger}=d_j$ and $c_j^{\dagger}=c_j$ (we choose a normalization such that $\{c_j,c_j\}=2$ and similarly for $d_j$). The Hamiltonian (1) then becomes :
\begin{equation}
H=-i\sum_{j=2m-1} (J_1 c_jd_{j+1}-J_2d_{j+1}c_{j+2}).
\end{equation}

\begin{figure}[t]
    \centering
      \includegraphics[scale=1]{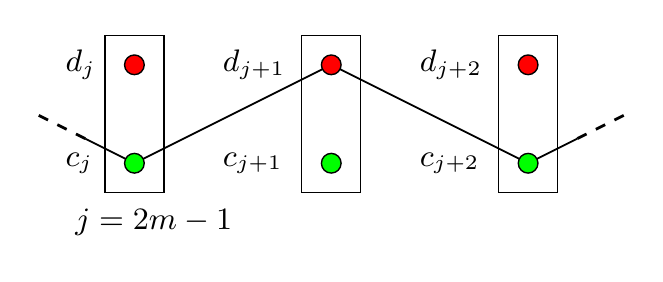} 
          \includegraphics[scale=1]{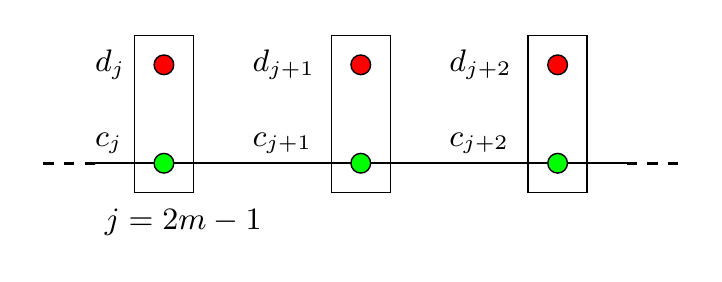} 
  \vskip -0.5cm  \includegraphics[scale=1.2]{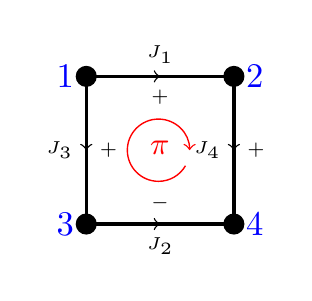} 
 \caption{(color online) Top: Majorana fermion configuration in Eq. (7). Middle: Alternative representation of Majorana fermions in Eq. (9). 
Bottom:  In ladder geometries discussed in Sec. III, one can define $\mathbf{Z}_2$ gauge fields by analogy with the Kitaev spin model \cite{Kitaev06}; see the Kitaev Hamiltonian in Eq. (19) and a comparison with our ladder Hamiltonian in Eq. (17). These gauge fields are represented by + and - choices of the $u_{j,k}$ variables of a ladder geometry. The gauge fields that take $+$ and $-$ values satisfy the condition that $u_{j,k}$ (to go from $j$ to $k$) is equal to $-u_{k,j}$; see the Kitaev Hamiltonian in Eq. (19).  The couplings $J_3$ and $J_4$ in Sec. III correspond to ferromagnetic couplings along the $Z$ direction. Fixing the gauge configurations for vertical bonds will also fix the parity operators for the $d$ Majorana fermions on these vertical bonds, and therefore the loop operator $D_{1,3}D_{2,4}$ defined in Sec. IIIC will be fixed to $+1$ or $-1$.}
 \label{majorana_1D}
\end{figure}

Note that, within these definitions the Majorana operators $\{c_{2m},d_{2m-1}\}$ are ``free'' (see fig.~\ref{majorana_1D} top). Alternatively, we can define the Majorana fermions as
\begin{eqnarray}
      c_j &=& i(a_j^{\dagger}-a_j) \hskip 0.3cm  d_j=a_j^{\dagger}+a_j \hskip 0.3cm   j=2m-1 \nonumber \\
      c_j &=& a_j^{\dagger}+a_j \hskip 0.3cm d_j=i(a_j^{\dagger}-a_j) \hskip 0.3cm  j=2m.		
    \end{eqnarray}
The Hamiltonian (1) becomes:
\begin{equation}
H=-i\sum_{j=2m-1} (J_1 c_j c_{j+1}-J_2c_{j+1}c_{j+2}).
\end{equation}
This Hamiltonian leads to the Majorana representation of Fig. 4 (middle). In the $A_x$ phase, we must satisfy for the ground state $-i c_{2m-1} c_{2m}=+1$ in (10) which is equivalent to $-i c_{2m-1} d_{2m}=+1$ in (8). Note that formally in the ground state $\sigma_{2m-1}^x \sigma_{2m}^x = - i c_{2m-1} c_{2m} = +1$ after the transformation (9).  It is also relevant to emphasize the difference of sign in front of the couplings $J_1$ and $J_2$; this will lead to $+$ and $-$ $\mathbf{Z}_2$ gauge fields discussed below for ladder systems with the definitions in Fig. 3.

The $d_j$ Majorana particles are now decoupled on each site, as illustrated in Fig. 3 (middle).  For the ground state, we note that $[H,i d_j d_{j+1}]=0$. For two successive sites, $i d_j d_{j+1}=\pm 1$ (meaning that spin correlation functions along $y$ direction on a given bond in the $A_x$ phase have equal probabilities to be $+1$ or $-1$). Then, we recover the $2^M$ quantum degeneracy of the chain due to bond formation in the spin liquid phase $A_x$. In Appendix B, we also analyze the braiding protocole of two Majorana fermions $d_j$ and $d_{j+1}$ with $j=2m-1$. The protocol and measurement are done in the spin space. We shall emphasize that in a single chain architecture, the d-Majorana fermions are not protected against noisy local magnetic fields. We note recent proposals in cQED to control the parity operator $p$ (between these two sites defined in Appendix B) \cite{Eran} and measure Majorana pairs \cite{Matthieu}. It is also relevant to mention recent progress in circuit quantum electrodynamics to measure spin observables and correlation functions \cite{Wallraff2} as well as in ultra-cold atoms \cite{Greiner2}.  In particular, Ref. \onlinecite{Zhong} reports the observation of a $\pi$ phase due to braiding and anyon statistics in a loop system of four qubit sites. In Sec. III. C, after discussing the phase diagram of ladder systems, we will discuss in more details the possibility to build loop qubit operators in relation with Fig. 3. Several theoretical proposals have already suggested similar architectures (in higher dimensional spaces) to engineer Majorana (code) constructions \cite{Terhal,Fu,AltlandEgger,Flensberg}. This is also related to experimental progress in topological superconducting wire systems \cite{Marcus}. 

The Majorana representation of Fig. 3 (middle) also allows us to  study the spin-1/2 edge excitation in more detail, in the infinite time limit, when increasing the ratio $J_1/J_2$. More precisely, adding a coupling $J_1$ between the Majorana fermion $c_1$ and the Majorana fermion $c_2$ (which lives at energy $\pm |J_2|$), this can produce virtual excitations shifting the $c_1$ Majorana fermion from zero energy. This results in a large but finite life-time for the spin-1/2 excitation at the edge, of the order of $|J_2|/J_1^2$. In this sense, the spin-1/2 excitation turns into a zero-energy $d_1$ Majorana fermion in the infinite time limit (which could also be sensitive to a local magnetic field along $X$ direction). A study of  such Majorana edge modes in inhomogeneous systems has been studied in Refs. \onlinecite{Smitha,Loss}. However, it is important to stress that in the $A_y$ gapped phase of the single chain, the spin-1/2 edge excitation is robust on time scales much longer than excitations in the bulk, which is in agreement with the spin analysis performed at the beginning of Sec. IIB. In addition, the winding number presented in Appendix A is evaluated directly on the Hamiltonian (3) in the Jordan-Wigner fermion basis and can equally reflect the presence of the spin-1/2 edge mode or of the $d_1$ Majorana fermion.

\section{Two-leg Ladders}

Now, we proceed with a detailed analysis of our phase diagram in Fig. 1. The boundary conditions and choice of parameters are adjusted to make the $A_x$ and $A_y$ phases symmetric here, i.e., with the same number of spin-1/2 edge excitations independently of the number of rungs). We note that the Square type ladder has been addressed in several works \cite{Feng,Wu,Smitha}, whereas the Brickwall ladder --- which is reminiscent of the honeycomb ribbon geometry --- has not been addressed so far, to the best of our knowledge. In the Brickwall ladder, we show that the gapless B phase of Fig. 1 is reduced to a line. This allows us to formulate an analogy to the occurrence of pre-formed pairs in the system. 

It is also important to mention exact constructions of chiral spin liquids \cite{Yao} and spin liquid states in ladder systems \cite{Meng}. Other exotic phenomena and Majorana edge modes have been addressed in Refs. \onlinecite{Motrunich,Loss}. Unusual phases can also appear in Majorana superconducting wire systems \cite{Loic,Rahmani} and in related hard-core boson ladders in relation with quantum Hall physics \cite{Alex,Leonardo,TeoKane}.  A discussion on symmetry protected topological phases has also been addressed in Ref. \onlinecite{Wenlad}. 

\subsection{The model}

We consider the spin $\frac{1}{2}$ system described in Fig. 4, with spins located on the vertices of two coupled chains of $2M$ sites each, with $l$ being the distance between two connected vertices (lattice spacing). The sites are labelled by two integers, the site index $j\in[\![1, 2M]\!]$ and the row index $\alpha\in\{1,2\}$. The Hamiltonian of the system reads :
\begin{equation}
H = H_1+H_2+H_I,
\end{equation}
where
\begin{eqnarray}
H_1&=&\sum_{j=2m-1}J_1 \sigma_{j,1}^x\sigma_{j+1,1}^x +J_2\sigma_{j+1,1}^y\sigma_{j+2,1}^y \nonumber \\
H_2 &=& \sum_{j=2m-1}J_1 \sigma_{j+1,2}^x\sigma_{j+2,2}^x +J_2\sigma_{j,2}^y\sigma_{j+1,2}^y \nonumber \\
H_I &=& \sum_{j=2m-1}J_3 \sigma_{j,1}^z\sigma_{j,2}^z + J_4 \sigma_{j+1,1}^z\sigma_{j+1,2}^z, 
\end{eqnarray} 
with $(J_1,J_2,J_3,J_4)<0$ being the coupling constants introduced in Fig. 1. The same formalism will allow us to treat in
a similar manner the Brickwall ladder characterized by  $J_4=0$, in fig.~\ref{fig_ladder_3}. Again, we note some invariance of the energy spectrum when changing $J_i \rightarrow -J_i$ simultaneously for all $i$.

\begin{figure}[t]
    \centering
      \includegraphics[scale=1.5]{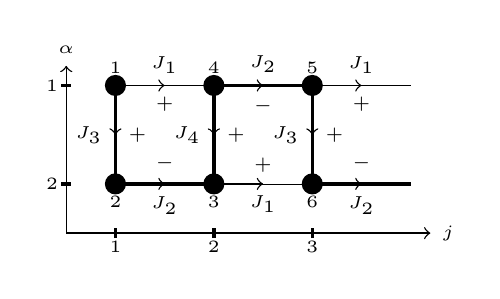} 
 \caption{Notations $(j,\alpha)$, where $j$ denotes the column and $\alpha$ the row, for the Square ladder. Gauge representation for $u_{i,j}$ and one string representation $1,2,3,4,...$ used for the Jordan-Wigner transformation (see Appendix C). Note that the definitions of sites $1, 2, 3, 4...$ is different than in Fig. 3, and agree with those in Appendix C.}
 \label{fig_ladder_4}
\end{figure} 

\begin{figure}[t]
	\centering
      \includegraphics[scale=1.5]{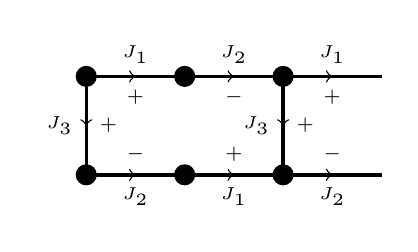}  
      \caption{Notations and Gauge choice in the case $J_4=0$ which corresponds to our brickwall or ribbon ladder.}
      \label{fig_ladder_3}
\end{figure}

\medskip

First, we write the Hamiltonian (12) in terms of fermionic operators, using the Jordan-Wigner transform. The Hamiltonian, which is obtained in Eq. (17), can be simply obtained using a Jordan-Wigner transformation for each chain separately as done in the previous Section. For completeness, in Appendix B, we show that the emergent Hamiltonian is independent of the chosen string path (one can choose a distinct path for the string operator.) For example, one can use the string (zig-zag) path of Fig. 4 (see Fig. 8b and 8c for different string configurations).  We follow the notations of Fig.~\ref{fig_ladder_4} and define

\begin{equation}
\left\{
\begin{array}{ll}
      \sigma_{j,\alpha}^- = a_{j,\alpha}e^{i\pi\sum_{\{i,\alpha\}\in string}a_{i,\alpha}^\dagger a_{i,\alpha}} \\
      \sigma_{j,\alpha}^+ = a_{j,\alpha}^\dagger e^{i\pi\sum_{\{i,\alpha\}\in string}a_{i,\alpha}^\dagger a_{i,\alpha}} 
    \end{array}
\right.
\hskip 0.3cm \alpha\in\{1,2\}.
\end{equation}

\begin{equation} 
\left\{
\begin{array}{ll}
      \sigma_{j,\alpha}^x =\sigma_{j,\alpha}^+ + \sigma_{j,\alpha}^- =  (a_{j,\alpha}^\dagger +a_{j,\alpha})e^{i\pi\sum_{i<j}a_{i,\alpha}^\dagger a_{i,\alpha}} \\
       \sigma_{j,\alpha}^y = \frac{1}{i}(\sigma_{j,\alpha}^+ - \sigma_{j,\alpha}^-) =  i(a_{j,\alpha}^\dagger -a_{j,\alpha})e^{i\pi\sum_{i<j}a_{i,\alpha}^\dagger a_{i,\alpha}}.
    \end{array}
\right.
\end{equation}
Furthermore, we introduce the Majorana fermions :
\begin{equation}
c_{j,\alpha}=\left\{
\begin{array}{ll}
      i(a_{j,\alpha}^\dagger-a_{j,\alpha}) \mbox{   ,   $j+\alpha=2m$}\\
      a_{j,\alpha}^\dagger+a_{j,\alpha} \mbox{   ,   $j+\alpha=2m-1$}
    \end{array}
\right.
\end{equation}

\begin{equation}
 d_{j,\alpha}=\left\{
\begin{array}{ll}
     a_{j,\alpha}^\dagger+a_{j,\alpha} \mbox{   ,   $j+\alpha=2m$}\\
      i(a_{j,\alpha}^\dagger-a_{j,\alpha}) \mbox{   ,   $j+\alpha=2m-1$}
    \end{array}
\right.
\end{equation}
In this construction, the Hamiltonian takes the form :
\begin{eqnarray}
H = -i \sum_{j=2m-1}  && [J_1c_{j,1}c_{j+1,1}-J_2c_{j+1,1}c_{j+2,1} \nonumber \\   
&+& J_1c_{j+1,2}c_{j+2,2}-J_2c_{j,2}c_{j+1,2} \\ \nonumber 
&+& J_3D_{j,1}c_{j,1}c_{j,2}+J_4D_{j+1,1}c_{j+1,1}c_{j+1,2}],
\end{eqnarray}
where $D_{j,\alpha}$ depends only on the $d_{j,\alpha}$ operators on a vertical bond through $D_{j,\alpha}=(-i)d_{j,\alpha} d_{j,\alpha+1}$. 

Here, $D_{j,\alpha}$ commutes with the Hamiltonian and can be seen as a classical variable in the ladder system which can be fixed in the ground state. More precisely, we can restrict the  study to the ("physical") subspace $S$, defined by : $|x\rangle \in S \iff D_{j,\alpha}|x\rangle = u_{j,\alpha}|x\rangle$ for all ${j,\alpha}$, with $u_{j,\alpha}=\pm 1$ an eigenvalue of $D_{j,\alpha}$ and where $S\subset S'$, $S'$ being the $4M$-dimensional Fock space in which the Majorana fermions live ("extended space").  We can then make a precise connection with the approach by Kitaev in two dimensions, where the spin operators in the expanded space are decomposed in terms of Majorana fermions $b^{\alpha}$ and $c_j$:
\begin{equation}
\sigma'^{\alpha}_j=ib^{\alpha}_jc_j \hskip 0.3cm \alpha\in\{x,y,z\}.
\end{equation}
Defining $u_{j,k}=ib_j^{\alpha}b_k^{\alpha}$ such that $\sigma'^{\alpha}_j \sigma'^{\alpha}_k = -i  u_{j,k} c_j c_k$, 
and re-labeling the sites using the string notation in Fig. 4, the Hamiltonian acquires the general form : 
\begin{equation}
H= \frac{-i}{2} \sum_{\langle j,k\rangle} J_{j,k} u_{j,k} c_j c_k,
\end{equation}
where the sum is performed over nearest neighbors $\langle j,k \rangle$. Within this notation $\langle j,k\rangle$ gives a factor 2 when summing over $j$ and $k$ since $u_{j,k}=-u_{k,j}$. The eigenvalues of $u_{j,k}$ are $u_{j,k}=\pm 1$ ; therefore, the variables $u_{j,k}$ can be seen as emergent $\mathbf{Z}_2$ gauge fields. (It is also important to mention that the $D$ operator defined above in terms of the $d$ Majorana fermions is not directly related to the other $D$ operator in the Kitaev paper \cite{Kitaev06}). Now, let us make an explicit connection with Eq. (17). 

With the string path chosen in Fig. 4, we extend the results of Ref. \onlinecite{Feng} to the case of general values of $J_3$ and $J_4$. It is important to note that the goal here is not to uniquely connect the Majorana basis $(d,c)$ introduced earlier with the Kitaev Majorana basis $(b_x,b_y,b_z,c)$ \cite{Kitaev06}. However, the $c$ fermions can be taken to be the same in Eq. (17) and Eq. (19). In addition, by comparing Eq. (17) and Eq. (19) then we can uniquely define the $u_{j,k}$ variables in the ladder.  It is already important to note that on horizontal links, in the Hamiltonian (19), the $u_{j,k}$ variables are just considered to be the pre-factors of the $J_1$ and $J_2$ terms, and therefore do not affect the parity operators $i d_{j,1} d_{j+1,1}$. From Eq. (19), indeed we see that the $u_{j,k}$ are already defined for the $J_1$ and $J_2$ links.

One needs now to fix the $u_{j,k}$ parameters on the $J_3$ and $J_4$ links. For this, we use Lieb's theorem \cite{Lieb} : in the case of the square ladder, the ground state of the system is in the $\pi$ flux configuration (and zero net flux if we consider two successive plaquettes) \cite{Feng}; to meet this requirement, we fix the $u_{j,k}$ to $+1$ for the vertical bonds, i.e. $D_{j,1}=D_{j+1,1}=1$ in (17), as illustrated in Fig. 4, in agreement with Lieb's theorem \cite{Lieb}.  We infer that the brick wall lattice model is in the zero flux ground state as a reminiscence of the two-dimensional model  (see Fig. 6) \cite{Kitaev06}. This flux choice allows us to bridge between the ribbon (honeycomb) and square ladders. 

The flux configurations alternate from $+$ to $-$ on a short length scale equal to the lattice spacing $l$, and must be treated exactly. The constrained choice on the gauge field $u_{j,k}$ now results on constraints for the $d$ Majorana fermions in a loop according to Fig. 4 (bottom). Note that we could have equally chosen a gauge with $D_{j,1}=D_{j+1,1}=-1$ in (17), and we will show below in Eq. (25) that the energy spectrum is invariant under the transformation $J_3\rightarrow -J_3$ and $J_4\rightarrow -J_4$. 

Re-injecting this choice of $\mathbf{Z}_2$ gauge fields in the Hamiltonian (17), we obtain the exactly solvable Hamiltonian:
\begin{equation}
 H=H_1+H_2+H_I
\end{equation}
with
$$H_1=-i\sum_{j=2m-1} (J_1c_{j,1} c_{j+1,1} - J_2 c_{j+1,1} c_{j+2,1}) \mbox{  ,  } m\in[\![1, M]\!]$$
$$H_2=-i\sum_{j=2m-1} (-J_2 c_{j,2} c_{j+1,2} + J_1 c_{j+1,2} c_{j+2,2})$$
$$H_I = -i\sum_{j=2m-1} (J_3 c_{j,1} c_{j,2} + J_4 c_{j+1,1} c_{j+1,2}).$$

\subsection{Energy spectrum}

In order to derive the spectrum of the above Hamiltonian, we note that the latter can be written in the general form :
\begin{equation}
H=-i\sum_{s\lambda, t\mu}J_{s\lambda,t\mu}c_{s\lambda}c_{t\mu},
\end{equation}
where, instead of labelling the sites with two integers, $(j,\alpha)$,  $j\in[\![1, M]\!]$, $\alpha\in\{1,2\}$, we changed the notation to $(s,\lambda)$, where $s\in[\![1, M]\!]$ denotes the cell index, and $\lambda\in[\![1, 4]\!]$ denoting the position of the site in a cell (see Fig. 4 bottom). Since $J_{s\lambda,t\mu}$ depends only on $s\lambda,t\mu$, the Fourier transform of the Hamiltonian then gives :
\begin{equation}
H=-i\sum_{k,\lambda, \mu}J_{\lambda,\mu}(k)c_{k,\lambda}c_{-k,\mu},
\end{equation}
with 
$$c_{k,\lambda}=\frac{1}{\sqrt{M}}\sum_{s=1}^{M} e^{-iksl}c_{s,\lambda}$$
and 
$$ J_{\lambda,\mu}(k)= e^{-i\textbf{k}\cdot(\textbf{r}_{s,\lambda}-\textbf{r}_{t,\mu})}J_{s\lambda,t\mu}.$$
We denote $\textbf{r}_{s,\lambda}$ the position of the site $(s,\lambda)$.

It is now straightforward to diagonalize this Hamiltonian (using the notations of Fig. 4 bottom) :
\begin{equation}
H=- i\sum_{k,\lambda,\mu} X^T\underbrace{\begin{pmatrix}
   0 & \alpha & \beta & 0 \\
   -\alpha^{*} & 0 & 0 & \gamma \\ 
   -\beta^{*} & 0 & 0 & -\alpha^{*}\\
   0 & -\gamma^{*} & \alpha & 0
\end{pmatrix}}_{M}X
\end{equation}
with $\alpha=J_1e^{-ikl}+J_2 e^{ikl}$, $\beta=J_3e^{-il}$, $\gamma=J_4e^{-il}$ and  
\begin{equation}
X=\begin{pmatrix}
 c_{-k,1}\\
 c_{-k,2} \\
 c_{-k,3} \\
 c_{-k,4}
\end{pmatrix}.
\end{equation}
The notations of the four fermions in a unit cell in k-space are chosen to recover block-diagonal matrices when $J_3=J_4=0$ (see Appendix A).

The energy spectrum of the Hamiltonian is given by the eigenvalues of the matrix :
\begin{equation}
\epsilon(k) = \pm \sqrt{\frac{A(k)}{2}\pm\frac{\sqrt{A(k)^2-4B(k)}}{2}},
\end{equation}
with 
$$A(k)=2\left(J_1^2+J_2^2+2J_1J_2\cos(2kl) + \frac{J_3^2+J_4^2}{2}\right)$$ 
and 
$$
\begin{array}{ll}B(k)=(J_1^2+J_2^2+2J_1J_2\cos(2kl))^2\\
+2J_3J_4(2J_1J_2+(J_1^2+J_2^2)\cos(2kl))+J_3^2J_4^2.
\end{array}$$

We shall now study the phase diagram of the system. Given a quadruplet
$(J_1,J_2,J_3,J_4)$, the spectrum of the corresponding Hamiltonian is gapless if there exists a mode $k$ such that $\epsilon(k)=0$. Thus, we need to find for which set of values of the coupling parameters the equation $\epsilon(k)=0$ has a solution. Note that : $\epsilon(k)=0$ is equivalent to $B(k)=0$. This equality results in the location of the gapless phase $B$ in the phase diagram of Fig. 1, for the generalized ladder with distinct $J_3$ and $J_4$ couplings. We also insist on the fact that the gapless $B$ phase is reduced to two transition lines for the Square ladder studied in Ref. \onlinecite{Feng}, as we also reproduce. We give some physical understanding of the emergence of such gapless excitations in Sec. IVA (along the gapless line of Fig. 6).

\subsection{Phase diagram of Fig. 1 and Known Limits}

First, let us check known limits in the ladder. Fixing $J_3=J_4=0$, in each chain, first we check the results of Sec. IIA. For the Square ladder with $J_3=J_4$, we recover the phase diagram of Ref. \onlinecite{Feng}. 
The choice of the $\pi$ flux configuration for the Square ladder can be understood as follows. First, note that a zero flux configuration would change the sign in front of the first term in $J_3 J_4$ in the second line of the definition of $B(k)$. The two choices of flux configuration would approximately give the same ground state energy at large $J_3=J_4$. Now, if we set $J_2=0$ for example, then the system will select the flux configuration such that $B(k)$ is minimum and therefore the ground state energy will be minimum. Since single chain systems exhibit excitations at $\cos(2 kl)=-1$ corresponding to flip a spin-1/2 in the $A_x$ phase, then the $\pi$ flux configuration will be favored. Since $J_3=J_4$, all vertical bonds are then identical. If we would have chosen a symmetric choice for the two chains in terms of $J_1$ and $J_2$, we would obtain instead a zero flux configuration in agreement with Ref. \onlinecite{Motrunich}. 

In addition to the Kitaev spin liquid phases $A_x$ and $A_y$ characterized by an intra-chain pairing contribution similar to Eq. (3), we also note the emergence of an $A_z$ phase, where the fermions now pair between chains favoring $|++\rangle=|+_1 +_2 \rangle_z$ and $|--\rangle=|-_1 -_2\rangle_z$ states polarized along the $z$ axis; $1$ and $2$ refer to the two chains and $+$ and $-$ correspond to the two spin eigenvalues of the spin-1/2 on each site with a polarization along $z$ axis. This $A_z$ spin liquid phase is also characterized by very short-range correlation functions (each vertical bond selects its own ground state configuration for the two spins) and a large quantum degeneracy in the ground state $2^{2M}$ (even for finite $J_1$ and $J_2$ in the spin language). More precisely, let us set $J_1=J_2=0$ in Eq. (17), such that each rung is formally decoupled from the neighboring rungs. Then, on a given rung we must satisfy $\sigma_{j,1}^z \sigma_{j,2}^z = D_{j,1}(-i c_{j,1} c_{j,2})=+1$ in the ground state; therefore formally in Eq. (17), there is a hidden double degeneracy on each rung reproducing the two states $|++\rangle=|+_1 +_2 \rangle_z$ and $|--\rangle=|-_1 -_2\rangle_z$. More precisely, we can also write $\sigma_{j,1}^z \sigma_{j,2}^z = -(i c_{j,1} d_{j,1})(i c_{j,2} d_{j,2})$. Changing $\sigma_{j,1}^z\rightarrow - \sigma_{j,1}^z$ formally means changing $ i c_{j,1} d_{j,1} \rightarrow  -i c_{j,1} d_{j,1}$.  In Sec. IVA, we will explicitly use the fact that the ground state at large $J_3$ has a large quantum degeneracy and that the ground state is a tensor product state. 

Based on Fig. 4, we also expect that far in the $A_x$ and $A_y$ phases, the system still exhibits one Jordan-Wigner fermion $a$ localized at each edge of the ladder and corresponding to gapless spin-1/2 excitations. The edge modes turn into Majorana excitations in the infinite time limit when $J_2$ or $J_1$ become finite, in agreement with Refs. \onlinecite{Smitha,Loss}; see Sec. IIB. In the two phases, the two edge modes appear on the same chain at the two extremities. In Appendix A, we present the winding number 
 for this situation following Ref. \onlinecite{Wu}. In the $A_z$ phase, the Majorana fermions are all paired (gapped). In this case, we do not expect gapless spin-1/2 excitations at the edges of the ladder. Similarly, the $B$ phase (or transition lines) can be described by resonating vertical bonds, as shown in Sec. IV, and therefore should not support gapless edge excitations. We check this point in Appendix A at a quantum critical point where the winding number becomes zero at the phase transition. 

The emergence of the gapless $B$ phase in the generalized phase diagram of Fig. 1 can also be understood from a dual mapping, using the notations of Ref. \onlinecite{Feng}. The Hamiltonian can be indeed re-written as \cite{Dutta}
\begin{eqnarray}
H = \sum_j J_1 \tau_{2j-2}^x \tau_{2j}^x - J_2 \tau_{2j-2}^y \tau_{2j}^y + J_3 \tau_{4j}^z + J_4 \tau_{4j+2}^z.
\end{eqnarray}
The fact that the odd sites do not enter in the mapping reflects the macroscopic degeneracies of the different phases. Using the change of variable $2j-2\rightarrow j-1$ and $2j\rightarrow j$, then we obtain a spin-1/2 XY chain with alternating transverse fields. The Hamiltonian
is solved exactly using the Jordan-Wigner transformation of Sec. IIA and one recovers a gapless spectrum when $J_3 J_4=(J_1-J_2)^2$ which corresponds to $B(k)=0$. 

Now, we discuss in more details our phase diagram of Fig. 6 obtained for the brickwall or ribbon ladder
$(J_4=0)$. When fixing the condition $J_4=0$ in Eq. (25) corresponding to the Brickwall ladder or honeycomb ribbon, we find that there is a transition line characterized by $\epsilon(k)=0$ and therefore by gapless excitations for all $J_3$ when $J_1=J_2$ (when fixing the condition 
$\cos(kl)=0$ or $\cos(2kl)=-1$ in the $4\times 4$ matrix). The system is always gapped for $J_1\neq J_2$. In Sec. IVB and Appendix A, we show that excitations along the gapless line can be in fact re-written as superpositions of Majorana fermions, resulting
in a low-energy fixed point which can be represented as gapless electron and hole excitations and a $U(1)$ Luttinger theory. In this basis, we note a small shift of the chemical potential such that excitations are slightly moved from the condition $\cos(2kl)=-1$.
Note that the limit $J_1=0$ and $J_2=0$ yields 4 degenerate levels on a rung not coupled with a $J_3$ coupling.

It is also relevant to note that by fixing $J_2=0$ and $J_1=-\infty$, the energy spectrum remains gapped for all values of $J_3$ and there is no quantum phase transition. We check this point explicitly in Appendix D computing spin correlation functions. 

\subsection{Majorana Loop Qubit}

Before studying in more detail the line of quantum phase transition found in the brick wall ladder in Fig. 6 (see Sec. IV), we address the possibility to realize qubit loop or plaquette operators encoded in the Majorana variables and showing some protection due to the emergent ${\cal Z}_2$ symmetry. For simplicity, we consider the generalized ladder system of Fig. 1 in the $A_z$ phase.

Following Kitaev \cite{Kitaev06} (and the notations of Sec. IIIA),  in a given loop of four sites, the emergent $\mathbf{Z}_2$ gauge fields depicted in Fig. 3 will be fixed to $+$ or $-$. The exact configuration will be fixed in agreement with Lieb's theorem \cite{Lieb}. For the Square ladder, with the choice of spin couplings in Fig. 4, the ground state will be in a $\pi$ flux configuration \cite{Feng} (meaning that the product of gauge fields $u_{j,k}$ defined in (19) will be $-$). The vertical bonds will exhibit the same gauge flux $+$ (or equivalently $-$). This will imply that the operators (on vertical links) $D_{1,3}=(-i)d_{1} d_{3}$ and $D_{2,4} = (-i)d_{2} d_{4}$ are fixed to the same value $+1$ or $-1$ in the ground state. 

In the loop composing the unit cell of a ladder in Fig. 3, then one can introduce a four-spin operator $\sigma_1^z \sigma_2^z \sigma_3^z \sigma_4^z$. Other possible plaquette operators have been discussed in Ref. \onlinecite{Loss}. Suppose now that we focus on the $A_z$ phase of the two-leg ladder system in Fig. 1, such that the fermions $c_i$ are all gapped and the product $c_1 c_2 c_3 c_4$ then is fixed to $+1$ or $-1$ in the ground state. One can then define the reduced ${\cal Z}_2$ Majorana qubit definition  ${\cal P} =  d_1 d_2 d_3 d_4$, 
which is fixed to $+1$ in the ground state since it is formally equal to $D_{1,3}D_{2,4}$, and ${\cal P} {\cal P}^{\dagger}= {\cal P} {\cal P} = 1$. Based on the discussion of Sec. IIIA, we note that ${\cal P}=+1$ still allows $i d_1 d_2=\pm 1$ and $i d_3 d_4=\mp 1$.  Formally, this conclusion indeed implies that the two chains are entangled and therefore that one focusses on the $A_z$ phase (in the $A_x$ and $A_y$ phases the Majorana fermions entangle in each chain separately). Now, let us discuss the braiding operation of $d_1$ and $d_2$ by adding 
a coupling $\delta J_2$ (see Appendix B).  By braiding the two Majorana fermions $d_1$ and $d_2$ (by changing $\delta J_2\rightarrow -\delta J_2$ on a link coupled with a $J_1$ coupling) then mathematically ${\cal P} \rightarrow - {\cal P}$ and therefore this loop operator has eigenvalues $+1$ and $-1$, by analogy to a qubit. Note that by braiding $d_1$ and $d_2$, formally $D_{1,3}\rightarrow D_{2,3}$ and $D_{2,4}\rightarrow D_{1,4}$. This exemplifies how to activate and measure such a qubit in the original spin language. This ${\cal P}$ operator can be seen as an analogue of the plaquette excitation operator in the toric code.

In the ladder architecture, this suggests that the operator ${\cal P}$ is also protected against small noisy magnetic fields (smaller than the energy scale associated to the gap of the Majorana fermions in the $A_z$ phase). As a result of the $\pi$ flux configuration for the ground state, both the $c$ and $d$ fermions are paired. One could measure the correlation function of $\sigma_{1}^z \sigma_{2}^z$ and $\sigma_{1}^y \sigma_{2}^y$  to detect the state of the qubit ${\cal P}$ after braiding of the Majorana fermions $d_1$ and $d_2$ (see Appendix B). The coupling $\delta J_2$ must be smaller than the energy gap  (to protect the structure of the ground state). 

\begin{figure*}[t]
\centering
\includegraphics[width=0.45\textwidth]{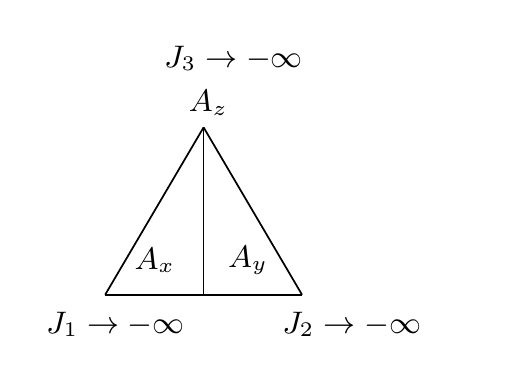} \hskip -0.3cm
\includegraphics[width=0.52\textwidth]{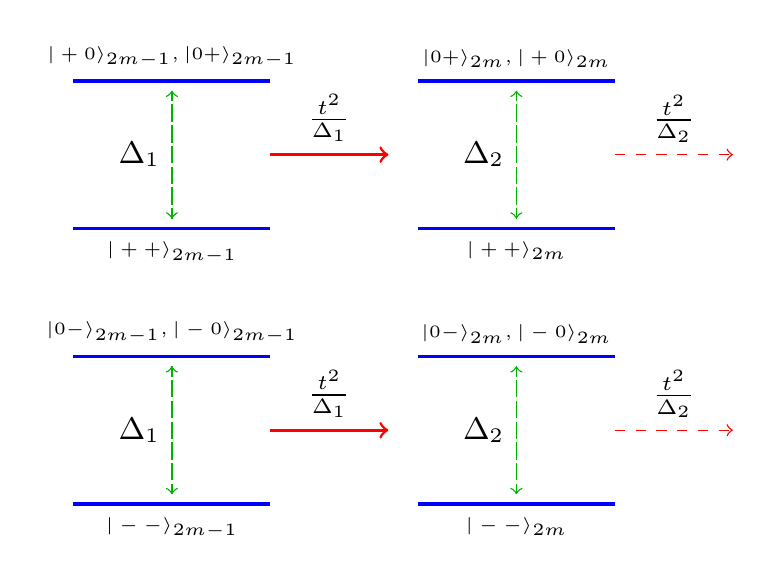}
\caption{(color online) (left) Our phase diagram in the case $J_4=0$. The gapped spin-liquid phase $A_x$ exhibits a polarization on a strong link that adiabatically passes from $x$ to $z$ direction along one side of the triangle by fixing $J_2=0$ and increasing $|J_3|$ (see Appendix D). Similarly, for the other vertical side of the triangle, starting from the $A_y$ phase, the spin polarization progressively passes from $y$ to $z$ direction. The vertical black line traduces the emergence of gapless spin excitations when $\cos(kl)=0$ or $\cos(2 kl)=-1$ which are studied in Sec. IV from perturbation theory and from bosonization. (right) Second-order Perturbation theory representation for the propagation of a pair $|++\rangle$ and $|--\rangle$ (or inversely a hole pair) in Sec. V from the rung $2m-1$ to $2m$ in the intermediate regime of $J_3$ where $\Delta_1$  and $\Delta_2$ are comparable, but still we can allow a small asymmetry between these two gap parameters. Formally, such a small asymmetry creates a Peierls type instability in the bulk (opening a gap). }
\label{phase_diag_J4eq0.pdf}
\end{figure*} 

These Majorana spin chains could offer a platform to realize artificial $\mathbf{Z}_2$ gauge fields on a lattice, and produce quantum gates applied directly on the Majorana basis. In a macroscopic system composed of several loops (ladders or two-dimensional lattice), the ground state satisfies conservation laws such as the conservation of total parity operator (product of all plaquette operators). One can then build multi-plaquette excitations. An example of possible operations for the honeycomb lattice is given in Ref. \onlinecite{Fu}, by analogy to the toric code \cite{toriccode,Wencode,Belen}. It is important to underline the experimental progress in cQED and Josephson junctions to implement related geometries \cite{Martinis,Zhong,Wallraff,Fowler,Benoit}, as well as in ultra-cold atoms \cite{atomcode}, to test braiding mechanisms and anyon statistics. A prototype device with Rydberg atoms has been studied in Ref. \onlinecite{Weimer}. The effect of defects have also been addressed, for example, in Refs. \onlinecite{Jiang,Petrova}. 

Brickwall ladders studied below could allow us to construct similar plaquette operators with $6$ sites as in Ref. \onlinecite{Fu}.

\section{Line of Gapless Spin Excitations}

In this Section, we study in more detail the two-leg ladder of Fig. 5 along the line of gapless spin excitations (identified for the ribbon Kitaev ladder or Brickwall ladder in Fig. 6). We assume therefore that $J_1\sim J_2$. 

First, in the regime of large $|J_3|$, we perform a perturbation theory in $J_1$ and $J_2$ showing how the states $|+_1 +_2\rangle_z=|++\rangle$, $|-_1 -_2\rangle_z=|--\rangle$ now can reside in the ground state on each bond, and resonate along the chains. Again, these states are defined with a quantization along the $Z$ direction since we start at the top of the triangle in Fig. 6. Then, we show that the propagation of such gapless spin excitations will be reinforced at small $| J_3 |$ based on a bosonization approach. This is also consistent with the Majorana description of Appendix A (ladder section). The bosonization approach also confirms that such gapless excitations become gapped for $J_1\neq J_2$. 

\subsection{Perturbation Theory}

First, we consider Fig. 6 on the gapless line for $J_1=J_2=0$ and $J_3\rightarrow -\infty$. The spin-spin correlation functions decay exponentially,  $\langle\sigma_j^z\sigma_{j+k}^z\rangle\propto e^{-\frac{|r_j-r_{j+k}|}{\xi}}$ with $\xi\propto |J_3|^{-1}\sim l$, indicating the emergence of a rung tensor product states (or matrix product states) representation \cite{Cirac}, by analogy with the $A_z$ phase of the Square ladder. At site $j=2m-1$, a possible state $|+_1+_2\rangle$ or  $|-_1-_2\rangle$ belonging to the ground state $|GS\rangle$ (which is fixed by the $J_3$ coupling) does not affect the spin polarization at the next rung labelled as $j+1=2m$.  For $J_1=J_2=0$ the ground state involves states of the form
\begin{equation}
|\mu\mu\rangle_{2m-1} \otimes |\nu\nu'\rangle_{2m},
\end{equation}
and the variables $\mu$, $\nu$ and $\nu'$ can take values $+$ or $-$ on a given cell of two successive rungs. Below, we show that the ground state remains of the same form after applying a small perturbation in $J_1$ 
and $J_2$. 

More precisely, let us start with a state $|++\rangle_j$ at the rung $j$ and we could consider different initial states of the form $|\nu\nu'\rangle_{2m}$ at the next rung. Let us apply a perturbation theory in $J_1 J_2$ where the process $J_1$ occurs first, for example (the order of operations does not affect the result), in analogy to the Kramers-Anderson magnetic induced coupling. The intermediate state involves an excited state with energy 
$2|J_3|$ from the ground state; this corresponds to flip one spin on a strong link. We obtain the following final state configurations:

$$
\begin{array}{ll}
J_2(\sigma_{j,2}^y)(\sigma_{j+1,2}^y) \frac{1}{2|J_3|}J_1(\sigma_{j,1}^x)(\sigma_{j+1,1}^x) |++\rangle_j \otimes
\left|
\begin{array}{ll}
|++\rangle_{j+1}\\
|--\rangle_{j+1}\\
|+-\rangle_{j+1}\\
|-+\rangle_{j+1}
\end{array}
\right.
\\
= \frac{J_1J_2}{2|J_3|} |--\rangle_j\otimes 
\left|
\begin{array}{ll}
-|--\rangle_{j+1}\\
|++\rangle_{j+1}\\
|-+\rangle_{j+1}\\
-|+-\rangle_{j+1}.
\end{array}
\right.
\end{array}
$$

Similarly, if we now consider an initial state  $|--\rangle_j$ at the rung $j$: 

$$
\begin{array}{ll}
J_2(\sigma_{j,2}^y)(\sigma_{j+1,2}^y)\frac{1}{2|J_3|} J_1(\sigma_{j,1}^x)(\sigma_{j+1,1}^x) |--\rangle_j\otimes 
\left|
\begin{array}{ll}
|++\rangle_{j+1}\\
|--\rangle_{j+1}\\
|+-\rangle_{j+1}\\
|-+\rangle_{j+1}
\end{array}
\right.
\\
= \frac{J_1 J_2} {2|J_3|} |++\rangle_j\otimes 
\left|
\begin{array}{ll}
|--\rangle_{j+1}\\
-|++\rangle_{j+1}\\
-|-+\rangle_{j+1}\\
|+-\rangle_{j+1}.
\end{array}
\right.
\end{array}
$$

Essentially, the four states $|--\rangle$, $|++\rangle$, $|-+\rangle$ and $|+-\rangle$ could now lie in the ground state if we sum over all possible choices of rungs and configurations.
Thus, the ground state remains of the same general form.  This implies that the correlation function $\langle\sigma_j^z\sigma_{j+k}^z\rangle$ still decays exponentially, with a characteristic length of the order of the lattice spacing. From this analysis, we also deduce that a state $|++\rangle$ on a rung $j$ can now propagate (in the Hilbert space of the ground state) to the successive rungs, and similarly for the state $|--\rangle$ assuming $J_1\sim J_2$.  In contrast, in the $A_z$ phase of the Square ladder, by applying say a $J_1$ coupling on a given chain, then the system would immediately react through the $J_3=J_4$ vertical couplings to restore the magnetic ground state (and the system is gapped). 

We could also include in our discussion fourth order contributions in perturbation theory and discuss the propagation of these states on successive rungs. More precisely, let us consider the state $|--\rangle_{j=2m-1} \otimes | --\rangle_{j=2m}$ obtained after the second-order perturbation theory. Now, let us consider the coupling between the state $ | --\rangle_{j=2m}$ and the successive rung $j=2m+1$ which involves two spins coupled by the strong-coupling $J_3$. This rung can be preferably in the state $ | --\rangle_{j=2m+1}$ or $ | ++\rangle_{j=2m+1}$. If we consider the state $ | --\rangle_{j=2m+1}$ and apply the perturbation theory in $J_1 J_2$ another time, then the state $| --\rangle_{j=2m} \otimes |--\rangle_{j=2m+1}$ will be changed into $|++\rangle_{j=2m} \otimes | ++\rangle_{j=2m+1}$. The ground state of these 3-rungs then will turn into  $|--\rangle_{j=2m-1} \otimes |++\rangle_{j=2m} \otimes | ++\rangle_{j=2m+1}$, exemplifying the propagation of gapless excitations along the chains. 

The introduction of one hole will increase the magnetic energy by $\Delta_1\sim | J_3 |$ on a rung $j=2m-1$ and by $\Delta_2 \sim J_1 J_2/(2| J_3 |) $ in a rung $j=2m$, as shown in Fig. 6. For two holes, then it will be preferable that they pair to minimize  the cost in magnetic energy.

\subsection{Pre-formed Pairs from Bosonization}

Based on the Majorana approach of Sec. III, we note that for the Brickwall ladder, there is a novel quantum phase transition line with gapless excitations for all values of $J_3$. To describe this point analytically for small values of $J_3$, we now apply the bosonization approach \cite{Haldane,Giamarchi}. We also address a connection with the Square ladder.

Below, we keep the choices that $J_i<0$, such that $J_3<0$ favors an attraction between the effective Jordan-Wigner fermions in the two chains introduced in Sec. IIA and III. This approach is then useful to see the appearance of pre-formed pairs of charge $2e$ in the model. More precisely, the two chains can be seen as $\uparrow$ $(+)$ and $\downarrow$ $(-)$ fermionic degrees of freedom of a Hubbard model coupled with an attractive interactions, making an analogy with a Luther-Emery liquid \cite{LutherEmery}. 

We can start from $H=(H_1+H_2)$ in Eq. (12) as two uncoupled chains with $J_1=J_2$. The terms $H_1$ and $H_2$ then take the same form as in Sec. IIA :
$$H_1=2J_1\sum_k \cos(kl)a_{k,1}^{\dagger}a_{k,1}$$
and by symmetry 
$$H_2=2J_1\sum_k \cos(kl)a_{k,2}^{\dagger}a_{k,2}.$$
The Jordan-Wigner fermions $a_{k,1}$ and $a_{k,2}$ are associated with the two chains (in each chain, we use the transformation (2) individually, for $J_3=0$). 

We now turn on the coupling $J_3$ :
\begin{eqnarray}
\lefteqn{H_I=J_3\sum_{j=2m-1}(1-2 a_{j,1}^{\dagger} a_{j,1})(1-2a_{j,2}^{\dagger}a_{j,2})}\\ \nonumber
&=& 4J_3\sum_{j=2m-1}a_{j,1}^{\dagger}a_{j,1}a_{j,2}^{\dagger}a_{j,2} \\ \nonumber
&-& \delta\mu\sum_{j=2m-1}\left(a_{j,1}^{\dagger}a_{j,1}+a_{j,2}^{\dagger}a_{j,2} - \frac{1}{2}\right),
\end{eqnarray}
where the chemical potential is renormalized to $\delta\mu=2J_3$. This perturbative theory in $J_3$ is thus valid as long as $\delta\mu\ll |J_1+J_2|$, such that the fermions maintain a linear spectrum in Fig. 2 (on the lattice, using the Majorana approach above, this seems to suggest that gapless excitations occur for $\cos(2kl)=-1$. In Appendix A, we suggest a change of basis trying to describe this small chemical potential shift and the fixed point below).  The effect of the small variation of the chemical potential $\delta \mu$ will not affect the low-energy fixed point, described below.

Now, we can apply bosonization in each chain $\alpha=(1,2)$ and use a continuum description where $a_{j,\alpha}$ is replaced by $a_{\alpha}(x)$. We introduce the left and right-moving electron fields around each Fermi point (Fig. 2) and relate in a standard way Fermi operators as exponential functions of bosonic operators $\theta_{\alpha}(x)$ and $\phi_{\alpha}(x)$ in each chain  \cite{Haldane,Giamarchi}. The 
(particle) density operator in each chain then takes the form \cite{Haldane,Giamarchi}:
\begin{eqnarray}
a^{\dagger}_{\alpha}(x)a_{\alpha}(x) &=& -\frac{\partial_x\phi_{\alpha}}{\pi} \\ \nonumber
&+& e^{-2ik_Fx}\frac{e^{-i(\theta_{\alpha}(x)+\phi_{\alpha}(x))}e^{i(\theta_{\alpha}(x)-\phi_{\alpha}(x))}}{2\pi l} \\ \nonumber
&+&e^{2ik_Fx}\frac{e^{i(\theta_{\alpha}(x)+\phi_{\alpha}(x))}e^{-i(\theta_{\alpha}(x)-\phi_{\alpha}(x))}}{2\pi l}.
\end{eqnarray}
We have the standard commutation relations: $[\phi_{\alpha}(x),\theta_{\alpha'}(x')]=i\frac{\pi}{2}\delta_{\alpha\alpha'}\mbox{Sign}(x-x')$. Then, we obtain the following Hamiltonian $H=H_1+H_2+H_I$:
\begin{eqnarray}
\label{boso}
H &=&\sum_{\alpha=1,2}\frac{v}{2\pi}\int dx ((\nabla\phi_{\alpha}(x))^2+(\nabla\theta_{\alpha}(x))^2) \\ \nonumber
&+&\int dx (f_{12}\nabla\phi_1(x)\nabla\phi_2(x)-\frac{b_{12}}{l^2}\cos(2(\phi_1(x)-\phi_2(x))),
\end{eqnarray}
with $v\sim -J_1l>0$, $f_{12}=\frac{J_3l}{\pi^2}<0$ and $b_{12}=-\frac{J_3l}{2\pi^2}>0$. The definitions of the bare parameters are adjusted such that we have an effective lattice spacing equal to $2l\rightarrow 0$. We note that umklapp scatterings involving $4k_F$ processes are not relevant here (first, due to the sign of $J_3<0$ and second due to the small shift of the chemical potential $\delta \mu$).

The term $f_{12}\nabla\phi_1(x)\nabla\phi_2(x)$ corresponds to forward scattering contributions and can be re-absorbed in the Gaussian contribution of Eq. (30) by a 
re-definition of the fields, as symmetric and anti-symmetric modes:
\begin{equation}
\phi_{\pm}=\frac{\phi_1\pm\phi_2}{\sqrt{2}}\ \hbox{,}\ \theta_{\pm}=\frac{\theta_1\pm\theta_2}{\sqrt{2}}.
\end{equation}
The Gaussian contribution then takes the form
\begin{eqnarray}
H_g &=& \frac{v_+}{2\pi}\int dx \frac{1}{K_+}(\partial_x \phi_+(x))^2+K_+(\partial_x \theta_+(x))^2 \\ \nonumber
&+& \frac{v_-}{2\pi}\int dx \frac{1}{K_-}(\partial_x\phi_-(x))^2+K_-(\partial_x\theta_-(x))^2
\end{eqnarray}
with 
$$
\left\{
\begin{array}{ll}
v_+K_+=v\\
\frac{v_{+}}{2\pi K_{+}}=\frac{v}{2\pi}+\frac{f_{12}}{2}
\end{array}
\right.
\left\{
\begin{array}{ll}
v_-K_-=v\\
\frac{v_{-}}{2\pi K_{-}}=\frac{v}{2\pi}-\frac{f_{12}}{2},
\end{array}
\right.
$$
and therefore $$K_{\pm}=\sqrt{\frac{v}{v\pm f_{12}/2}}.$$
Note that : $K_{+}>1$, $K_-<1$. In the sense of conformal field theory, such a theory described by $H_g$ would be described by a central charge $c=2$ \cite{Tsvelik}, referring to two gapless $U(1)$ theories. However,  the scattering term $b_{12}$ in Eq. (\ref{boso}) can open a gap in the sense of the renormalization group arguments. In fact, denoting the Hamiltonian as $H=H_g+H_{b_{12}}$ we find that such a term grows under renormalization group arguments at large length scales or low energy (see Appendix E), and therefore the 
 ground state will pin the field $\phi_{-}$ to one of the classical minima of the cosine potential, opening a mass term 
 \begin{equation}
m^{*}\sim |J_1|\left(\frac{J_3}{J_1}\right)^{\frac{1}{2-2K_-}}. 
\end{equation}
Through this continuum description, the gap becomes equal on each bond since formally we have taken $2l\rightarrow 0$. This is also in agreement with the strong-coupling approach of Sec. IVA which suggests that the gaps on nearest-neighbor rungs would become equal when decreasing the inter-chain coupling. The model (with central charge $c=1$) nevertheless remains gapless because the mode $\phi_{+}$ remains gapless \cite{Tsvelik}. Note that the anti-symmetric mode $\phi_-(x)$ refers to high-energy (gapped) excitations associated with states $|+ - \rangle$ and $|- +\rangle$ in the spin language. The ground state at small $J_3$ allows gapless spin excitations in the (symmetric) sector $|++\rangle$ and $|--\rangle$. In the fermionic or bosonized representation, we recover that these gapless spin excitations refer to the propagation of charge $2e$ (Cooper pairs between chains) in the system, associated with the symmetric mode $\phi_+(x)$. We also confirm that here the sign of $J_3$ matters. The opening of the mass term assumes that $K_-<1$ and therefore that $J_3<0$, or attractive interactions between fermionic chains.  In addition, $K_+>1$ usually refers to attractive interactions in one dimension. The emergent fixed point then shows some analogy with a Luther-Emery liquid if we identify the mode $\phi_-(x)$ as a spin degree of freedom (or relative charge density) \cite{LutherEmery}. 

From the bosonization theory, we can compute spin-spin correlation functions and we obtain (see Appendix E):
\begin{equation}
\langle\sigma_{\alpha}^{x,y}(x)\sigma_{\alpha}^{x,y}(0)\rangle\propto e^{-|x|/\xi},
\end{equation}
with $\xi\propto \frac{1}{m^*}$. In the $z$ direction,  we find a power-law decay of the correlation function (as a reminiscence of the situation at $J_3=0$):
\begin{equation}
\langle\sigma_{\alpha}^{z}(x)\sigma_{\alpha}^z(0)\rangle\propto \frac{K_+}{x^2}+\frac{(-1)^{x/l}}{x^{K_+}}.
\end{equation}
We thus have an algebraic spin liquid. If we include the role of a small chemical potential shift $\delta \mu$ in the discussion, this does not modify the conclusion; indeed, the chemical potential shift $\delta \mu$ involves the symmetric mode $\phi_+$. As long as the energy spectrum in Fig. 2 remains linear, we infer that the velocity of the mode $\phi_+$ would remain unchanged and that all the results remain identical. These results can also be in principle checked using the Majorana approach by recombining fermions, as discussed at the end of Appendix A. Increasing the ferromagnetic coupling $|J_3|$ we observe that on the one hand, the states $|+ - \rangle$ and $|- +\rangle$ acquire a larger gap and on the other hand, the Luttinger parameter $K_+$ increases meaning that the system will converge more and more to a matrix product states representation (or short-range spin liquid).  The system will exhibit equally strong ferromagnetic and anti-ferromagnetic correlation functions along the chain direction. Formally, we observe that $K_+$ diverges when $v+f_{12}/2=0$, indicating that the method is not valid anymore for $J_1\sim J_3/(2\pi^2)$. 

In addition, it is important to notice that the mass term $m^*$ appears due to a coupling between four fermions, which is (almost) a marginal coupling at small $J_3$. Therefore, we deduce that as soon as we deviate from the symmetric condition 
$J_1=J_2$, then the intra-chain BCS terms in Eq. (3) will become more important, emphasizing the fact that the physics described in this Section is valid at the quantum phase transition only between the two phases $A_x$ and $A_y$.

To summarize, related to Fig. 6, we confirm that the system is fully gapped for $J_1\neq J_2$ (spin liquid phase). For $J_1=J_2$ the system exhibits gapless excitations which can be seen as analogues of preformed charges $2e$ propagating along the chains. This approach complements then the efforts at large $|J_3|$ presented above.

\subsection{Square ladders and bosonization}

Before addressing the case of doping the system with a pair of holes, we briefly make an analogy with Square ladders. 

Based on the Majorana approach of Sec. IIIB, we observe that  the Square ladder is described by two distinct transition lines characterized by the condition $|J_1-J_2| = |J_3|$, in accordance with Ref. \onlinecite{Feng}.  This is also in agreement with Fig. 1 if one sets $J_3=J_4$. We identify two transition lines separating the $A_z$ phase from the $A_x$ phase, and the $A_z$ phase from the $A_y$ phase.
If $|J_3|$ is small enough (compared to $|J_1|$ and $|J_2|$), one can address the physics along these lines in a similar manner as Eq. (30) by considering a small asymmetry in the velocities $v_1$ and $v_2$ associated with the two modes $\phi_1$ and $\phi_2$ (or equivalently associated with the two modes $\theta_1$ and $\theta_2$). The rest of the description is unchanged since the bosonization continuum description assumes a vanishing lattice spacing and therefore the forms of $b_{12}$ and $f_{12}$ remain the same. Such a small asymmetry in the velocities, gives a coupling of the form $\nabla \phi_+(x) \nabla \phi_-(x)$ at the fixed point and similarly $\nabla \theta_+(x)\nabla \theta_-(x)$. As long as $|J_3|$ is sufficiently small compared to the energy gap $m^*$, then classically one can approximate $\nabla \phi_-\approx 0$. In addition, $\nabla \theta_-$ becomes an irrelevant operator in the sense of the renormalization group and therefore should not affect (deeply) the fixed point. Nevertheless, proceeding along the lines of Ref. \onlinecite{Urs}, then one can integrate out the antisymmetric mode $-$ at the fixed point exactly. This would only renormalize the Luttinger parameter $K_+$. 

We then conclude that the transition lines in the Square ladder at small $J_3$ could also be described by a similar Luttinger theory. Starting from the $A_z$ phase of the square ladder then the system would yield pre-formed pairs $|++\rangle$ and $|--\rangle$ becoming gapless towards the transitions with the $A_x$ and $A_y$ phases. A connection between bosonized quantum field theories and Ising transitions has also been noted in Refs. \onlinecite{Loic,Kane} in different models.

\section{A hole pair in the Mott state}

Here, we study the effect of a few holes in the system starting from the gapless line of the Brickwall ladder. This allows us to start with the rung product state representation of Sec. IVA. We build a perturbative analysis in Sec. VA and B. We show the possibility to observe
an insulating-superconducting transition for hole pairs in the dilute limit. The insulating phase is topological in the sense that hole pairs will localize at the boundary. The emergence of the superconducting phase can be intuited from the bosonization approach where we have already identified pre-formed pairs. These preformed pairs also occur in the $A_z$ phase of the Square ladder (see Sec. IVC). Therefore, the arguments of Sec. VC below are also applicable to the slightly doped $A_z$ phase.

To study the motion of holes in the dilute limit, we proceed as follows. First, we assume that the magnetic ground state $|GS\rangle$ at large $|J_3|$ (related to Eq. (27)) containing pre-formed pairs is not modified by the addition of a few holes. This means that the propagation of a few holes will be treated perturbatively, modifying only weakly the total energy of the system. Using the properties of Mott phases that spin-1/2 magnetizations are associated with electron spins, we introduce the electron creation and annihilation operators :
$$c_{j,\mu}^{\alpha \dagger}, c_{j,\mu}^{\alpha}.$$
These operators respectively create and annihilate a spin in the state $\mu=\uparrow,\downarrow$ on the site $j$ of the chain $\alpha=1,2$. The hopping of a hole from a site $j+1$ to the site $j$ is thus described by :
\begin{equation}
c_{j+1,\mu}^{\alpha \dagger}c_{j,\mu}^{\alpha}|\mu\rangle_j\otimes|0\rangle_{j+1}=|0\rangle_j\otimes|\mu\rangle_{j+1}.
\end{equation}
 We assume that double occupancy is suppressed on each site as a result of a large on-site Hubbard repulsion, which has produced the Mott phase, with one (localized) electron per site at half-filling (or one spin-1/2 per site). The Hamiltonian describing the motion of a hole along each chain then takes the form :
\begin{equation}
H_{1h}=-t\sum_{j; \mu=\uparrow,\downarrow; \alpha=1,2}c_{j+1,\mu}^{\alpha \dagger}c_{j,\mu}^{\alpha}+h.c. , 
\end{equation}
where $t$ is the (effective) hopping amplitude along the chains. (In the regime studied below, two holes on a given rung will pair and therefore we do not need to introduce inter-chain hopping of single
holes \cite{Heinz,KarynMaurice}.)

\begin{figure*}[t] 
    \centering
       \includegraphics[scale=1.4]{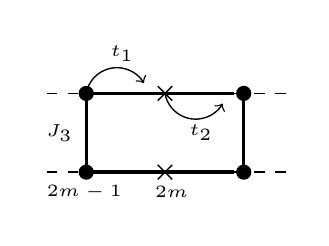}   \vskip -0.2cm
         \includegraphics[scale=1.3]{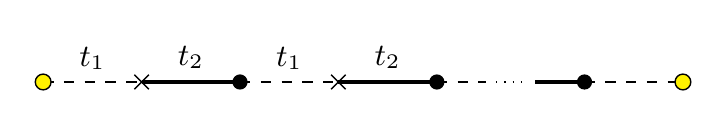}    
       \caption{(color online) Top: Effective hopping amplitudes when exchanging a magnetic bond $|++\rangle$ and $|--\rangle$ and a hole pair. 
       Here, $t_1\sim t^2/\Delta_1$ and $t_2\sim t^2/\Delta_2$ denote the effective hopping amplitudes obtained from second-order 
       perturbation theory in the intermediate region of $J_3$ for the Brickwall lattice at $J_1=J_2$. We stress that this approach is not valid at large $J_3$ and holes would occupy rungs with crosses.
       Bottom: Mapping to an effective SSH model; the yellow states denote two quasi-zero energy states for the hole pair when $t_1<t_2$. For a finite size system, the overlap between the two edge wave functions will produce symmetric and
       anti-symmetric combinations of the edge excitations. The system becomes analogous to a topological insulator with a charge gap in the bulk and hole-pair excitations prepared at the edges in an adiabatic manner at time $t=0$ \cite{Bryce}. When $J_1=J_2\sim \sqrt{2}J_3$, we predict a phase transition when $t_1=t_2$ at small $J_3$, which corresponds to a quasi-one-dimensional superconductor: the hole pair does not feel the effect of the boundary at time $t>0$ and coherently propagates in the system.}
\end{figure*}

We work in the dilute limit following Ref. \onlinecite{Heinz}, and therefore we neglect the hole correlation functions on different sites \cite{Maurice}. Formally, in the limit of $|J_3|\rightarrow +\infty$, the hole will preferably localize on a rung $j=2m$ in Fig. 7 to minimize the magnetic exchange. Indeed, the introduction of a hole is equivalent to suppress a spin-1/2 particle on a rung and therefore would increase the energy by an amount $|J_3|$ on a rung $j=2m-1$ in Fig. 7.  Propagation of single holes have been addressed theoretically in two dimensions \cite{Chalker,Moessner}. Here, for very large $|J_3|$, single hole motion could also occur to second-order in perturbation theory in $t$. To be more precise, we expect that single-hole physics will be important when $J_1/J_3\rightarrow 0$ and $J_2/J_3\rightarrow 0$ and $t>(|J_1|, |J_2|)$. To second-order perturbation theory in $t$, a single hole could then preferably tunnel from the rung $j=2m$ to the rung $j=2m+2$ or the rung $j=2m-2$.

Below, we address in contrast the propagation of two holes (2 spin vacancies) or even number of holes in the dilute limit where $J_1=J_2$ is not so distinct from $J_3$. This corresponds to situations with intermediate values of $J_3$.

\subsection{Hole Pair Propagation}

More precisely, we consider situations where the ground state (magnetic) energy of a pair $|\mu \mu\rangle$ on a rung $j=2m-1$ (which is equal to $J_3$) and that of the same pair on a rung $j=2m$ (which is equal to $-J_1 J_2/(2|J_3|)$) are not so distinct such that holes can occupy the two rungs of the ladder; second-order energy corrections in $t$ found below will compensate for the energy difference $\Delta_1-\Delta_2$ defined in Sec. IV.A. 

In this limit, to minimize the magnetic energy (see Sec. IVA), it is then favorable for two holes to form a pair on a rung $j$ where one hole localizes on the site $j$ of each chain. This reduces the number of  affected magnetic bonds and connects with the
(coherent) propagation of states $|++\rangle$ and $|--\rangle$ in the system. Let us consider a protocol similar to Fig. 6 right. A hole pair is at the rung $j+1=2m$ in the initial state (again, we assume that the system is sufficiently long such that the rest of the
system remains in the same magnetic ground state).  Let us consider the motion of a state $|++\rangle$ or $|--\rangle$ from a rung $j=2m-1$ to $j+1=2m$ .  The initial state on this cell is $|\mu \mu\rangle_{2m-1}\otimes|00\rangle_{2m}$. The exchange between the hole pair and the magnetic bond  $|++\rangle$ or $|--\rangle$ then is described by the process :
\begin{equation}
-t_{1}c_{j+1,\mu}^{2\dagger}c^{1\dagger}_{j+1,\mu}c_{j,\mu}^{2}c_{j,\mu}^1|\mu \mu\rangle_{j}\otimes|00\rangle_{j+1}.
\end{equation}
Here, $t_1$ describes the effective hopping amplitude (for an illustration, see Fig. 7 top). We find $t_{1}=\frac{t^2}{\Delta_1}$ where $\Delta_1$ is roughly the energy cost to create a hole at site $j$ (see Fig. 6, right). 
Starting from the strong-coupling limit, $\Delta_1 \sim | J_3|$. 

Similarly, the hopping of a magnetic bond $|++\rangle$ or $|--\rangle$  from $j=2m$ to $j+1=2m+1$ is described by 
\begin{equation}
-t_{2}c_{j+1,\mu}^{2\dagger}c^{1\dagger}_{j+1,\mu}c_{j,\mu}^{2}c_{j,\mu}^1|\mu\mu\rangle_{j}\otimes|00\rangle_{j+1},
\end{equation}
with $t_{2}=\frac{t^2}{\Delta_2}$ and starting from the strong-coupling regime we estimate $\Delta_2\sim \frac{J_1J_2}{2|J_3|}$ (see Fig. 7 top and Fig. 6). We also note from the bosonization study of Sec. IVB, that in the intermediate
regime of $J_3$, the gaps $\Delta_1$ and $\Delta_2$ should not be too different (at small $J_3$, formally the two gaps are equal to $m^*$). Note that the emergence of asymmetric values of $\Delta_1$ and $\Delta_2$ for the intermediate regime of $J_3$ is not in contradiction with the bosonization study of Sec. IVB, which stops to be valid at $J_1\sim J_3/(2\pi^2)$ and therefore cannot reveal a situation where $\Delta_1\neq \Delta_2$ (due to the continuum limit restriction). 

It is important to underline that we consider the situation where $t_2\ll \Delta_2$ and $\Delta_1$ is close to $\Delta_2$, meaning that $t_1$ and $t_2$ are not too distinct. Formally, in this perturbation theory, we neglect the modification of the ground state energy by an amount $\Delta_1 - \Delta_2$. Therefore, this scheme is applicable when second-order energy corrections in $t_1$ and $t_2$ are larger than $(\Delta_1 - \Delta_2)$, or $\Delta_1(\Delta_1-\Delta_2) <t^2\ll \Delta_2^2$. We assume that these conditions are fulfilled below. The case $t_1<t_2$ can be realized corresponding to intermediate values of $J_3$ and the case $t_1=t_2$ corresponds to smaller values of $J_3$ in agreement with the bosonization approach where $\Delta_1=\Delta_2=m^*$. Using the strong-coupling forms of $\Delta_1$ and $\Delta_2$ related to Sec. IVA, the condition $t_1=t_2$ or $\Delta_1=\Delta_2$ occurs for $J_1=J_2\sim \sqrt{2}J_3$ in the intermediate regime of $J_3$.  

\subsection{SSH model for the hole pair}

We can now introduce the bosonic operators corresponding to the creation/annihilation operator of a hole pair such that:
\begin{equation}
\left\{
\begin{array}{ll}
a^{\dagger}_j=c_{j,\mu}^2c_{j,\mu}^1\\
a_j=(c_{j,\mu}^2 c_{j,\mu}^1)^{\dagger} 
\end{array}
\right.
\mbox{, $j=2m-1$ },
\end{equation}
and
\begin{equation}
\left\{
\begin{array}{ll}
b^{\dagger}_j=c_{j,\mu}^2c_{j,\mu}^1\\
b_j=(c_{j,\mu}^2 c_{j,\mu}^1)^{\dagger}
\end{array}
\right.
 \mbox{, $j=2m$},
\end{equation}
where $\mu=\uparrow$ or $\downarrow$. Formally, we introduce a unique hole pair operator (independent of the flavor $\mu$ because the hopping of a hole pair in one direction is equivalent to the hopping of a pair $|+ +\rangle$
or $|--\rangle$ in the opposite direction; see Fig. 6).

Since we do not allow more than one electron per site, the bosons $a_j$ and $b_j$ are in fact hard core bosons or spins. We can apply the Jordan-Wigner transformation on these operators and re-write by analogy to the slightly doped $t-J-J_{\perp}$ ladder \cite{Heinz} 
the Hamiltonian as a one-dimensional tight-binding model of spinless fermions.
Then, the motion of a hole pair is described by a spinless fermion Hamiltonian with two inequivalent sites $a$ and $b$ denoted by crosses and circles (and $M$ unit cells separated from a distance $l$) in Fig. 7 bottom:
\begin{equation}
H_{h.p.}= -t_{1}\sum_{j=2m-1} (a_{j}^{\dagger}b_{j+1}+h.c.) -t_{2}\sum_{j=2m} (b_{j}^{\dagger}a_{j+1}+h.c.).
\end{equation}
We note again a mapping towards the Su-Schrieffer-Heeger model introduced in polyacetylene \cite{SSH}. Going to the Fourier space, we define $a_j=\frac{1}{\sqrt{M}}\sum_k a_k e^{ikx_j}$
with $x_j=jl$ and $k\in ]-\frac{\pi}{l},\frac{\pi}{l}]$ of the form $k=\frac{2\pi p}{Ml}-\frac{\pi}{l}$, $p\in[\![1, M]\!]$. We use the notations of the $M$ unit cells with Fig. 7. We derive:
\begin{equation}
H_{h.p.}= \sum_k (a_k^{\dagger},b_k^{\dagger})
\begin{pmatrix}
   0 & h(k) \\
   h(k)^{*} & 0  
\end{pmatrix}
\begin{pmatrix}
a_k\\
b_k
\end{pmatrix},
\end{equation}
with $h(k)=-t_1 -t_2e^{ikl}$. The energy spectrum satisfies:
\begin{equation}
\epsilon(k)=\pm\sqrt{t_1^2+t_2^2+2t_1t_2\cos(kl)},
\end{equation}
by analogy to the case of the magnetic chain in Sec. II.  We check that the energy spectrum has a gap at $\cos(k l)=-1$ for $t_1\neq t_2$ meaning at the edges of the Brillouin zone for a one-dimensional tight-binding model. Formally, the chemical potential is equal to $\mu=0$ here and lies between the lowest and upper bands. Choosing a convention of unit cell where the gap occurs at $k = \pm \pi/l$ is physical for this analysis: on each rung, a pre-formed pair contributes to a `double' charge $2e$, therefore the occupancy of the effective Jordan-Wigner band  has doubled compared to the spin description in Fig. 2. The lowest band is then filled. We underline that Mott physics or infinite on-site repulsion has been taken into account by changing the statistics of hole pairs or pre-formed pairs from bosons to hard-core bosons or spinless fermions. 

Now, let us add a pair of holes with an energy equal to $\mu=0$ (between the valence and conduction band for $t_1<t_2$), at the edges at time $t=0$ (Fig. 7). To be prepared at zero energy, formally the coupling $t_1$ near the edges should be switched on adiabatically from zero. This process could maintain this additional hole pair (for a long time) at the edges, as shown experimentally in Ref. \onlinecite{Bryce} in ultra-cold systems.  For a finite size system, the hole-pair excitations would be symmetric and anti-symmetric combinations of the edge wave-functions \cite{Asboth,Carlos}. Following the notations of Ref. \onlinecite{Asboth}, we find a connection between the number of edge modes in a finite chain and the winding number of the bulk Hamiltonian \cite{Asboth}, by analogy to the $A_x$ and $A_y$ magnetic phases
of a single chain (see Appendix A): 
\begin{equation}
\nu = \frac{1}{2i\pi}\int_{-\pi/l}^{\pi/l} dk \frac{d\log{h(k)}}{dk},
\end{equation}
with $\log{h(k)}=\log{|h(k)|}+i\mbox{arg}(h(k))$. We expect to have edge modes in the case $t_1< t_2$, which would correspond to $\nu=1$, whereas for $t_1= t_2$ we expect the system to be non-topological, i.e. $\nu=0$.  For  $t_1< t_2$, we have $\log{h(k)}\approx \log{t_2}+i(kl+\pi)$, and we check $\nu=1$. Therefore, this implies that  similar to Fig. 7 (bottom), one could observe a hole pair with a $1/2$ probability to be localized on the left or on the right edge, producing edge excitations at zero energy \cite{Carlos}. A finite size system results in a small overlap between these two excitations; excitations become even and odd superpositions of these edge excitations. If the hole pair is added in a non-equilibrium manner (quench), still the probability to be at the edges at time $t$ can still reveal signatures of the boundary \cite{Bryce}. 

We check that $\nu=0$ for $t_1 = t_2$, implying a quantum phase transition in the system by decreasing $|J_3|$ until $J_1=J_2\sim \sqrt{2}J_3$ based on results of Sec. IVA (the bulk gap closes). The system will become similar to a quasi-one-dimensional superconductor (the pre-formed pairs discussed in Sec. IVB through $\phi_+$ can now leak in the system and are described by a Gaussian model); see Sec. VC. The case $t_1=t_2$ can also be realized in principle for $J_3=J_4$ in the Square ladder. This transition could be observed by decreasing $J_3$ along the line of gapless excitations in the brick wall lattice. Similar insulating-superconducting quantum phase transitions have been observed in Josephson junction arrays \cite{Chow,RMP}. 

We also underline that in principle we can still deviate slightly from the gapless line in the intermediate $J_3$ limit such that $|J_3|$ or the gap $m^*$ is larger than the energy scale $|J_1-J_2|$, which controls the occurrence of intra-chain pairing interaction of the Jordan-Wigner fermions in the spin-liquid phases $A_x$ and $A_y$. It is also important to underline that Thouless pump experiments in ultra-cold atoms have recently measured similar topological invariants \cite{Munich}. Such topological invariants in relation with quantum random walks has also attracted some attention recently \cite{Asboth2,Kitagawa}.

\subsection{Superconducting Transition}

Let us now consider the situation where $t_1=t_2$. This can be realized for the Brickwall ladder when $J_1=J_2\leq \sqrt{2}J_3$  or in the $A_z$ phase of the square ladder where $\Delta_1=\Delta_2=|J_3|$. The hole pairs can now coherently propagate in the system by analogy to a free fermion model (hard-core boson model) \cite{Heinz}. The system then could be seen as a superconducting spin liquid
\begin{equation}
|\Psi \rangle = |GS\rangle \otimes|\mbox{quasi-SC}\rangle,
\end{equation}
where the spin liquid part is described by the appropriate magnetic ground state $|GS\rangle$ (related to Sec. IVB at small $J_3$ or to the $A_z$ phase in Sec. IVC) and the hole pairs can propagate coherently forming a one-dimensional analogue of a superconductor (these hole pairs are described by a Luttinger theory with a Luttinger exponent equal to one in agreement with hard-core bosons and free spinless fermions \cite{Heinz,KarynMaurice}. This form of wave-function reproduces the emergent `spin-charge' separation (environmental magnetic RVB state and propagating hole pair) of ladder systems with a few hole pairs \cite{Heinz}. Using the analogy between hard-core bosons and free electrons in the Hamiltonian, we predict that the Green's function for the hole pairs then is given by :
\begin{equation}
\langle a_j^{\dagger} a_k\rangle \sim \frac{1}{|j-k|}.
\end{equation}
The hole pairs formed on a rung, defined in Eq. (41), have also p-wave symmetry. We note some analogy to the slightly-doped $SU(2)$ invariant spin ladder \cite{Heinz,KarynMaurice}. 

Coupling weakly identical spin ladders could result in long-range (superconducting) order  at zero temperature, by analogy to magnon excitations in weakly coupled spin ladders \cite{ThierryAlexei}. 

On related models, superconductivity was also predicted in two-dimensional doped Kitaev models \cite{Rosenow,Tianhan} and $t-J$ or related Hamiltonians
 \cite{Annica,Wei,Wen,Andrej,Chung-Hou,Doron}. A relation between resonating valence bond states, polyacetylene and superconductivity has been addressed in Ref. \onlinecite{Kivelson}.

\section{Conclusion}

To summarize, in this paper, we have studied networks of Kitaev magnetic chains and ladders. These systems can be engineered in superconducting quantum circuits  and ultra-cold atoms \cite{Martinischain,Duan} and are related to the discovery of quantum spin liquid materials \cite{Kee,Dima}.

Through the Jordan-Wigner transformation, this allows us to map $\mathbf{Z}_2$ quantum spin-liquid states with short-range interactions onto BCS quadratic superconducting Hamiltonians. The emergent p-wave symmetry associated with the pairing order parameter leads to the occurrence of Majorana particles in the system, then referring to Majorana RVB states. More precisely, the Kitaev spin chain in the bulk can be mapped onto a gapped p-wave superconductor plus a gapless chain of Majorana fermions. In ladder systems, the ground state selects a particular flux configuration allowing us to gap all the Majorana fermions by pairs. This loop device could be a first step to realize a Majorana code. This system could also bridge with implementations of the toric code to test anyon braiding statistics \cite{Zhong}, in the context of cQED. Some efforts have also been realized in ultra-cold atoms \cite{atomcode}. We have also underlined a connection between the Kitaev magnetic chain and the SSH model, from the emergence
of gapless excitations at the edges (both in the spin and Majorana representations).

In the Square ladder, based on a Majorana fermion representation, we have recovered an identical phase diagram as Ref. \onlinecite{Feng}, with the three gapped spin-liquid phases $A_x$, $A_y$ and $A_z$ by analogy with the two-dimensional Kitaev model. In the Brickwall ladder system, which corresponds to a ribbon geometry of the two-dimensional Kitaev honeycomb model, we have predicted a line of gapless (bulk) excitations in the phase diagram, connecting two spin-gapped phases. In a gapped phase, the spin polarization on a strong bond adiabatically bridges from $X$ to $Z$  or from $Y$ to $Z$. Along this line, based on perturbation theory and bosonization, we have identified gapless pre-formed pairs and excitations corresponding to the propagation of $|+_1+_2\rangle_z$ and $|-_1-_2\rangle_z$ (magnetic) states along the chains. We have revealed a magnetic analogue of a Luther-Emery liquid theory \cite{LutherEmery} with central charge $c=1$ in the limit of small $|J_3 |$. The two chains can be identified as a (pseudo) spin-up and spin-down degree of freedom in the Hubbard model, and are coupled through an attractive interaction since $J_3<0$.  We have also shown how the gapless line spreads out in a $B$ phase for the generalized ladder system.  

By doping the Brickwall ladder with a pair of holes, we have made another analogy with the symmetry protected topological SSH model, where a hole pair could localize at the edges with (almost) zero energy for intermediate values of $J_3$.  At small values of $J_3$, as a result of $|J_1|=|J_2|\gg |J_3|$, the pre-formed pairs on the strong vertical bonds propagate equally on all rungs. By doping with a pair of holes, the pre-formed pairs can propagate coherently along the chains producing a quasi-one-dimensional superconducting spin liquid. For the Square ladder, doping the $A_z$ phase, could also produce a similar superconducting spin liquid state with quasi-long-range order. This analysis reinforces the idea that the occurrence of superconductivity in ladder systems (and potentially in high-Tc superconductors) requires the formation of resonating valence bond states.
\\
\\
Acknowledgements: This work has benefitted from discussions with Eric Akkermans, F\' elicien Appas, Annica Black-Schaffer, Beno\^it Dou\c{c}ot, Tal Goren, Maria Hermanns, Lo\" ic Herviou, Tianhan Liu, Christophe Mora, St\' ephane Munier, Pascal Paganini, Belen Paredes, Kirill Plekhanov, Alexandru Petrescu, Stephan Rachel, Nicolas Regnault, C\' ecile Reppelin, T. Maurice Rice, Guillaume Roux, Ronny Thomale and Wei Wu. We also acknowledge discussions at KITP Santa-Barbara, Nordita and CIFAR meetings. We acknowledge funding from the Labex PALM Paris-Saclay and from the German DFG through a ForscherGruppe 2414. 

\appendix

\section{Fourier transform, Winding Number}

\subsection{Fourier Transform}

Here, it is instructive to focus on the spin chain and Fourier transform, to evaluate the winding numbers of the $A_x$, $A_y$ magnetic phases in relation with the SSH model.  These winding numbers encode the presence of additional excitations at the edges.
 
 First, let us start from Eq. (3) in the main text. We observe that in a given chain $J_1$ couples the combination $(a_j - a^{\dagger}_j)$ with the other combination  $(a_{j+1} + a^{\dagger}_{j+1})$ and similarly the coupling $J_2$ involves $(a_{j+1} - a^{\dagger}_{j+1})$ with the other combination  $(a_{j+2} + a^{\dagger}_{j+2})$. In Sec. IIB, we introduce a Majorana representation $c_j$ and $d_j$ to describe these distinct combinations at different sites. Formally, one can Fourier transform the Majorana operators $c_j$ and $d_j$, and introduce 
$c_k$ and $d_k$ such that $c_{-k}=c^{\dagger}_k$ and  $d_{-k}=d^{\dagger}_k$. Then, for a single chain we can write the Hamiltonian as:

\begin{equation}
H=-i\sum_{k} Y^T\underbrace{\begin{pmatrix}
   0 & \alpha  \\
   -\alpha^{*} & 0 \\
   \end{pmatrix}}_{M} Y
\end{equation}
with $\alpha= (J_1e^{-ikl}+J_2 e^{ikl})$ and 
\begin{equation}
Y=\begin{pmatrix}
 c_{-k,1}\\
 d_{-k,1} 
 \end{pmatrix}.
\end{equation}
(The subscript 1 refers to chain 1 in the ladder formulation.) Remember that in Eq. (10), the $d$ fermion is re-labelled $c$ to simply the notations in the ladder models. The energy eigenvalues are then equal to
\begin{equation}
\epsilon(k) = \pm \sqrt{| \alpha(k)|^2} = \pm \sqrt{J_1^2 + J_2^2 + 2 J_1 J_2\cos(2k l)}.
\end{equation}
The spectrum is invariant under the replacement $2k l \rightarrow 2k l + 2\pi$ with a reduced Brillouin zone $-\pi/(2l) \leq k \leq \pi/(2l)$. For the two chain systems, when $J_3=J_4=0$, the energy is additive and we check using the $4\times 4$ matrix representation of Sec. IIIB. We also use the fermionic basis $(a_k, a^{\dagger}_{-k})$ to check this result. For $J_1=J_2$, the gap closes and the low-energy spectrum close to $2k_F l=\pi$ takes the form $\epsilon(k) = 2J_1 \cos(k l)$ with $J_1<0$. 

\subsection{Gapless edge excitations}

We can introduce the Anderson pseudo-spin 1/2 representation or Pauli matrix representation related to Eq. (A1) \cite{Anderson}: the effective field reads $\alpha_1= \alpha= (J_1e^{-ikl}+J_2 e^{ikl})$, and can also be inferred from the $4\times 4$ matrix in the main text. Within this formulation, the Brillouin zone is reduced $-\pi/(2l) \leq k \leq \pi/(2l)$. The pseudo-spin encodes the information regarding the orientation of a strong bond, $X$ versus $Y$. To define the winding number from  $-\pi/l$ to $\pi/l$, we can reset the convention with  two inequivalent sites per unit cell where the cells are separated from $l$, and we find $\alpha_1(k) = (J_1 + J_2 e^{ilk})$ in a similar manner as the SSH model in Sec. V; the relative phase difference between the two terms is in agreement with the convention of the SSH model (which will be discussed in Sec. V) \cite{Asboth}. For a single chain model, the winding number
\begin{equation}
\nu = \frac{1}{2i \pi}\int_{-\pi/l}^{\pi/l} dk \left(\frac{d\log \alpha_1}{dk} \right),
\end{equation}
where $\log(\alpha_1) = \log |\alpha_1 | + i \arg(\alpha_1(k))$ is one or zero, similarly to the SSH model of Sec. V (reflecting the presence of edge Jordan-Wigner fermions by analogy to Fig. 7 when exchanging $t_i \leftrightarrow J_i$). More precisely, $\nu$ is one for $J_2>J_1$ and a even number of sites (or equivalently an integer number of unit cells) and zero when $J_1>J_2$. For $J_2>J_1$, we expect gapless excitations at the edges (as a reminiscence of the case $J_1=0$). In the single chain problem, the situation $J_2>J_1$ corresponds to the $A_y$ magnetic phase and $J_1>J_2$ corresponds to the $A_x$ magnetic phase. 

Note however that this calculation cannot distinguish between a spin-1/2 (or Jordan-Wigner fermion $a$, mixture of $c$ and $d$) and a Majorana edge excitation; for a discussion, see Sec. IIB. 

\subsection{Ladder Systems}

In the ladder systems, based on Fig. 4, we also expect one localized excitation at each edge of the ladder in the $A_x$ and $A_y$ phase, and no edge mode in the $A_z$ phase since the parity operators $D_{2j}$ and $D_{2j+1}$ are fixed in the ground state. The fact that the winding number in the $A_z$ phase is zero can be checked by plugging $J_1=J_2=0$ in Eq. (25) and we observe that the matrix elements do not depend on the wave-vector $k$. 

Below, we check in a simple manner the non-zero winding numbers for the $A_x$ and $A_y$ phases. The term $J_3$ (and or $J_4$) is assumed to be small. For the second chain, we should reverse the role of $J_1$ and $J_2$ based on our convention of Fig. 5, and we get the related pseudo-spin $\alpha_2(k)= (J_2 + J_1 e^{ikl})$. The related ket for the chain 2 is written as
\begin{equation}
Z=\begin{pmatrix}
 c_{-k,2}\\
 d_{-k,2} 
 \end{pmatrix}.
\end{equation}
(The subscript 2 now refers to chain 2.)
By analogy with the SSH model \cite{Asboth}, in the ladder system, we can then define the winding number as two additive contributions:
\begin{equation}
\nu = \frac{1}{2i \pi}\int_{-\pi/l}^{\pi/l} dk \left(\frac{d\log \alpha_1}{dk} +  \frac{d\log \alpha_2}{dk}\right),
\end{equation}
where $\log(\alpha_i) = \log |\alpha_i | + i \arg(\alpha_i(k))$. Formally, for $J_3=J_4=0$, we get two separate $2\times 2$ matrices. When $J_1\rightarrow 0$ or $J_2\rightarrow 0$, then this reduces to $\arg(\alpha_1)(k) + \arg(\alpha_2)(k) = ikl $. This implies $\nu=1$ when $J_1\neq J_2$. Essentially, this supports the idea that the phases $A_x$ and $A_y$ have one $a$ Jordan-Wigner fermion localized at each edge of the ladder, independently of the number of cells in the system. At the quantum critical point $J_1=J_2$, the spectrum of a given chain takes the form $2J_1\cos(kl)$ and therefore we obtain $\nu=0$. 

Now, let us describe the effect of $J_3$ in the brick wall ladder at the quantum phase transition. It is then instructive to define linear combinations of the Majorana fermions $c_{k,1}$ and $c_{k,2}$ in this $2\times 2$ matrix representation. We use that the $d$ fermions are gapped through the $D_j$ operators. Let us introduce $c_{k,1}^* = \frac{1}{\sqrt{2}}(c_{k,1} + i c_{k,2})$ such that $(c_{k,1}^*)^{\dagger} = \frac{1}{\sqrt{2}}(c_{-k,1} - i c_{-k,2})$. We observe that the $J_3$ coupling can be re-written in a diagonal manner as
\begin{equation}
(-i) J_3 c_{-k,1} c_{k,2} + h.c. = -2J_3 (c_{k,1}^*)^{\dagger} c_{k,1}^*.
\end{equation}
For the brick-wall ladder this argument tends to confirm that excitations for $J_1=J_2$ become described by electron and hole excitations in agreement with Sec. IVB. 

\section{Braiding Majorana fermions}

Here, we discuss the possibility to engineer braiding of Majorana fermions, using the description of Sec. II. This is an example of quantum operation described in the Majorana basis that can be activated and measured in the spin language.

Let us imagine that in Eqs. (1) and (10) we switch on a small coupling $\delta J_2\sigma_{2m-1}^y \sigma_{2m}^y$ on a link already coupled by a Ising coupling along $X$ direction.  The perturbation is smaller than the gap such that the $c$ fermions in the bulk are paired at higher energy. Then, this will induce a coupling $+ i \delta J_2 d_{2m-1} d_{2m}$. If $\delta J_2>0$ then, the ground state will select $i d_{2m-1} d_{2m}=-1$ and if $\delta J_2<0$, the ground state will select $i d_{2m-1} d_{2m}=+1$. Changing the sign of this coupling in time thus would correspond to a braiding of these two Majorana fermions in time. More precisely, the two ground states associated with this local exchange coupling $\delta J_2$ are related by exchanging $d_{2m-1} \leftrightarrow d_{2m}$ if we perform an operation $\delta J_2 \leftrightarrow - \delta J_2$. Doing such an operation is equivalent to control the parity operator for these two sites defined as $p=\exp (i \pi f^{\dagger} f)=(1-2f^{\dagger} f)$, with $f^{\dagger} f = i d_{2m} d_{2m-1} /2 + 1/2= 0\ \hbox{or}\  1$.   

The next question then is how to measure such a braiding process in the spin language? 
It is important to note that on the two site model $(2m-1,2m)$, when acting with the coupling $\delta J_2$,  the state $|+_{2m-1} +_{2m}\rangle_x$ becomes transformed into $(i)^2 |-_{2m-1} -_{2m}\rangle_x$, similarly to the double well. This reveals that in the spin language $\sigma_{2m-1}^x$ does not commute strictly with the Hamiltonian anymore and the ground state is a Einstein-Podolsky-Rosen composed of a superposition of states $|+_{2m-1} +_{2m}\rangle_x$ and $|-_{2m-1} -_{2m}\rangle_x$. In average, one finds $\langle \sigma_{2m-1}^x \sigma_{2m}^x \rangle = 1$ since $_x\langle  -_{2m-1} -_{2m} | \sigma_{2m-1}^x \sigma_{2m}^x  |-_{2m-1} -_{2m}\rangle_x = 1$; this correlation function is therefore not changed when changing the sign of $\delta J_2$. This will be taken into account below through $-i c_{2m-1} c_{2m}=+1$ by analogy with Eq. (10). Now, let us measure the correlations for these two sites along the $Z$ axis (the rest of the chain is assumed to be in the ground state of the $A_x$ phase). Since $[H, \sigma_{2m-1}^z \sigma_{2m}^z]=0$, $\sigma_{2m-1}^z \sigma_{2m}^z$ can be equal to $+1$ or $-1$.  More precisely, we can re-write  $\sigma_{2m-1}^z \sigma_{2m}^z$ in terms of $-c_{2m-1} d_{2m-1} d_{2m} c_{2m}$ (after the transformation (9)). Now, let us look at the evolution of this correlation function during the protocole where we change $\delta J_2$ into $-\delta J_2$. We suppose that $|\delta J_2|$ is changed adiabatically (and that $|\delta J_2|$ remains smaller than the gap of the $c$ fermions). We can then use the fact that  $-i c_{2m-1} c_{2m}=+1$. Therefore, changing the sign of $\delta J_2$ or braiding $d_{2m-1}$ and $d_{2m}$ would change the sign of the measured correlation function of $\sigma_{2m-1}^z \sigma_{2m}^z=-i d_{2m-1} d_{2m}$. In addition, this correlation function should remain quantized and equal to $+1$ or $-1$. One could also choose to measure correlation functions along the $Y$ direction, since we also have $\sigma_{2m-1}^y \sigma_{2m}^y = i d_{2m-1} d_{2m}$. By braiding $d_{2m-1}$ and $d_{2m}$, one would then observe a change of signs of these two correlation functions. 

This protocole (braiding and measurement with spins) will be applied in Sec. III. D for the ladder system. 

\section{Jordan-Wigner String for the ladder}

Here, we provide a simple derivation of the Hamiltonian (17) for two (other) distinct paths of the string operator associated with the Jordan-Wigner transformation. In particular, within such a choice, the coupling $J_1$ becomes highly non-local. 

\begin{figure}
\begin{center}
\includegraphics[width=0.5\textwidth]{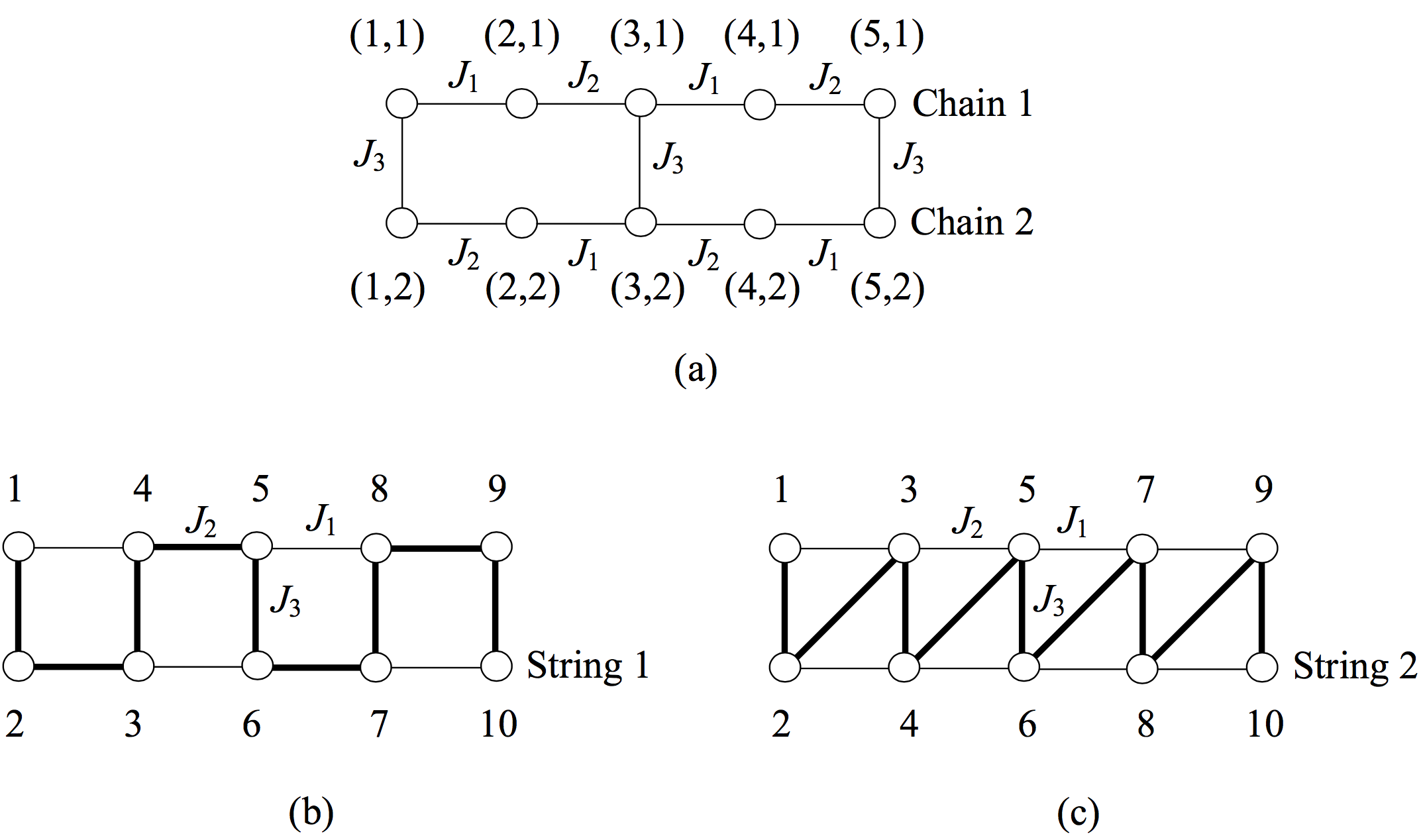}
\caption[]{(a) Indices $(j, l)$ denote $j$-th column and $l$-th row. (b) and (c) are two deformed string representations.} 
\label{fig:string}
\end{center}
\end{figure} 

The site $5$ is chosen as the reference site, and we derive the Hamiltonian coupling site $5$ with its neighbors. For String 1 in Fig.~\ref{fig:string} (b), the interaction from the $x$ direction on the $5$-th site reads
\begin{equation}
J_1\sigma_5^x\sigma_8^x = J_1( a_5^\dagger - a_5)(a_8^\dagger + a_8)e^{i\pi ( a_6^\dagger a_6 + a_7^\dagger a_7)}. 
\label{eq:j1}
 \end{equation}
As $[a_6^\dagger a_6 , J_1\sigma_5^x\sigma_8^x] = [ a_7^\dagger a_7, J_1\sigma_5^x\sigma_8^x] = 0$, there exists a common eigenspace in which the eigenvalues of $a_6^\dagger a_6$ and $a_7^\dagger a_7$ are $c$-numbers. We can choose a local gauge such that it minimizes total ground state energy of the string. So there should be no excited particle at sites 6 and 7, $a_6^\dagger a_6 = a_7^\dagger a_7 = 0$. Now we turn to Fig.~\ref{fig:string} (a) and define a set of Majorana operators, similarly as in the main text:
\begin{gather*}
c_{j,l} = \begin{cases} i(a_{j,l}^\dagger - a_{j,l}), & j + l = 2m; \\
a_{j,l}^\dagger + a_{j,l}, & j + l = 2m-1.
\end{cases}  \\ \
d_{j,l} = \begin{cases} a_{j,l}^\dagger + a_{j,l} , & j + l = 2m; \\
i(a_{j,l}^\dagger - a_{j,l}), & j + l = 2m-1. 
\end{cases}  
\end{gather*}
The Hamiltonian $(\ref{eq:j1})$ then can be transformed into
\begin{equation}
J_1\sigma_5^x\sigma_8^x = (-i) J_1c_{3,1} c_{4,1}. 
\end{equation}
Similarly, the other two interaction terms on 5-th site  become:
\begin{eqnarray}
J_2\sigma_4^y\sigma_5^y &=& -J_2( a_4^\dagger + a_4)(a_5^\dagger - a_5) = i J_2 c_{2,1}c_{3,1}, \nonumber \\
J_3\sigma_5^z\sigma_6^z &=& J_3(a_5^\dagger - a_5)(a_5^\dagger + a_5)(a_6^\dagger - a_6)(a_6^\dagger + a_6) \nonumber \\
&=& (-i) J_3 D_{3,1}c_{3,1}c_{3,2}, 
\end{eqnarray}
where we introduce the $D_{j,l}$ operators on vertical bonds: $ D_{j,l} = (-i)d_{j,l}d_{j,l+1}$.
Then, we get the Hamiltonian on the $5$-th site for the choice of String 1:
\begin{eqnarray}
H_{(5),\text{ String 1}}  &=& H_{(3,1),\text{ String 1}}  \\ 
&=& (-i) (J_1c_{3,1} c_{4,1} - J_2 c_{2,1}c_{3,1} + J_3 D_{3,1}c_{3,1}c_{3,2}). \nonumber
\end{eqnarray}

For the String 2 in Fig.~\ref{fig:string} (c), the interaction term on the $5$-th site involving $J_3$ remains the same. The $J_1$ and $J_2$ couplings then turn into:
\begin{eqnarray}
J_1\sigma_5^x\sigma_7^x &=&  J_1(a_5^\dagger + a_5)(1-2a_5^\dagger a_5)(a_7^\dagger + a_7)e^{i\pi a_6^\dagger a_6} \nonumber \\
&=& (-i) J_1c_{3,1} c_{4,1}, \nonumber \\
J_2\sigma_3^y\sigma_5^y 
&=& - J_2(a_3^\dagger - a_3)(1-2a_3^\dagger a_3)(a_5^\dagger - a_5)e^{i\pi a_4^\dagger a_4} \nonumber \\
&=& i J_2 c_{2,1}c_{3,1}, \nonumber
\end{eqnarray}
where $a_6^\dagger a_6 = a_4^\dagger a_4 = 0$; sites $4$ and $6$ are in the ground state of the string and there are no particle excitations. 
We confirm that the interaction Hamiltonian including the $5$-th site is the same for different routes of string: $H_{(5),\text{ String 2}}  = H_{(5),\text{ String 1}}$. 
We can generalize this approach to the full ladder and check Eq. (17). 

\section{Correlation functions in the honeycomb ribbon}

\begin{figure}
\centering
		\includegraphics[width=.9\linewidth]{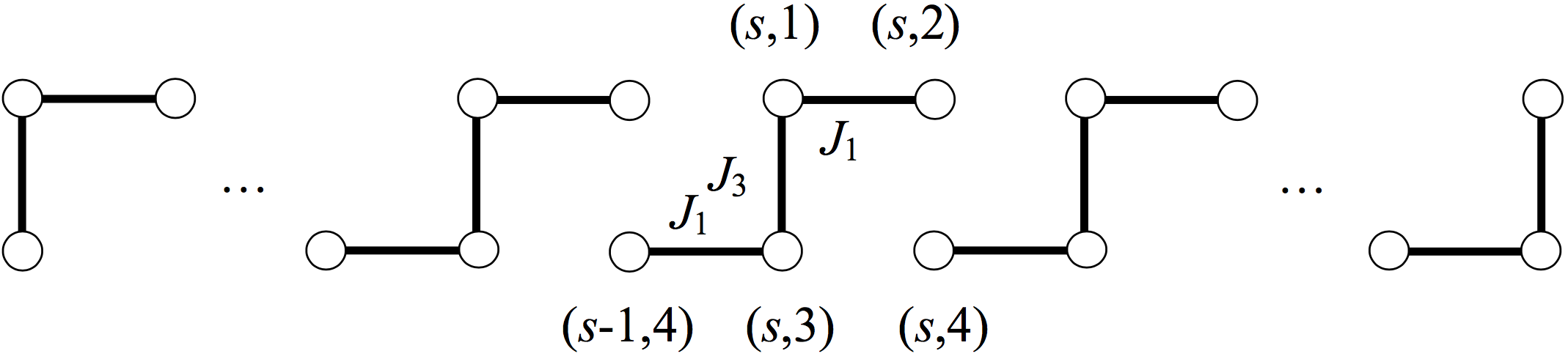} 
			\caption{Cluster honeycomb ladder.}
	\label{fig:ribbon}
\end{figure}

Here, we check the absence of quantum phase transition on the vertical sides of the triangle, for the ribbon phase diagram in Fig. 6. We check this fact explicitly by evaluating the spin correlation function(s) in a given unit cell, between nearest sites. For instance, when fixing $J_2=J_4= 0$, we get the cluster ladder of Fig.~\ref{fig:ribbon}.
The corresponding Hamiltonian matrix becomes
\begin{eqnarray}
H' = 2H &=& -i \sum_{k} X^T W X, \\ \nonumber
X^T &=& \begin{pmatrix} c_{k,1} & c_{k,2} & c_{k,3} & c_{k,4} \end{pmatrix}, \notag \\ \nonumber
W &=& \begin{pmatrix} 0 & \alpha & \beta & 0 \\ -\alpha^* & 0 & 0 & 0 \\ -\beta^* & 0 & 0 & -\alpha^* \\ 0 & 0 & \alpha & 0 \end{pmatrix}, 
\alpha = J_1e^{-ikl}, \beta = J_3 e^{-il}, 
\end{eqnarray} 
with the eigenvalues for the spectrum
\begin{equation}
\epsilon (k) = \pm \sqrt{J_1^2 + \frac{J_3^2}{2} \pm \frac{\sqrt{4J_1^2J_3^2 + J_3^4}}{2}}.
\end{equation}
We solve the eigenfunction for the ground state
\begin{eqnarray}
H' \left|GS\right> &=& \epsilon_0 \left| GS \right>, \\ \nonumber
\epsilon_0 &=& - \sqrt{J_1^2 + \frac{J_3^2}{2} + \frac{\sqrt{4J_1^2J_3^2 + J_3^4}}{2}}, \\ \nonumber
\left|GS\right> &=& \sum_k \frac{1}{\mathcal{N}} \left( x_1 c_{-k,1} + x_2 c_{-k,2}  + x_3 c_{-k,3} + x_4 c_{-k,4}\right) \left| 0 \right>, \\ \nonumber
x_1 &=& e, x_2 = -i\alpha^*, x_3 = \frac{e^2-\left| \alpha \right|^2}{i\beta}, x_4 = \frac{\alpha\left( e^2-\left| \alpha \right|^2 \right) }{e\beta}.
\end{eqnarray}
We introduce the variables
\begin{equation}
t = \frac{J_1}{J_3}, \qquad \frac{e}{\left| J_3 \right|} = \left(  t^2  + \frac{1}{2} \sqrt{4t^2+1}+ \frac{1}{2}\right)^{1/2}=g.
\end{equation}

\begin{figure}
\centering
	  \includegraphics[width=.7\linewidth]{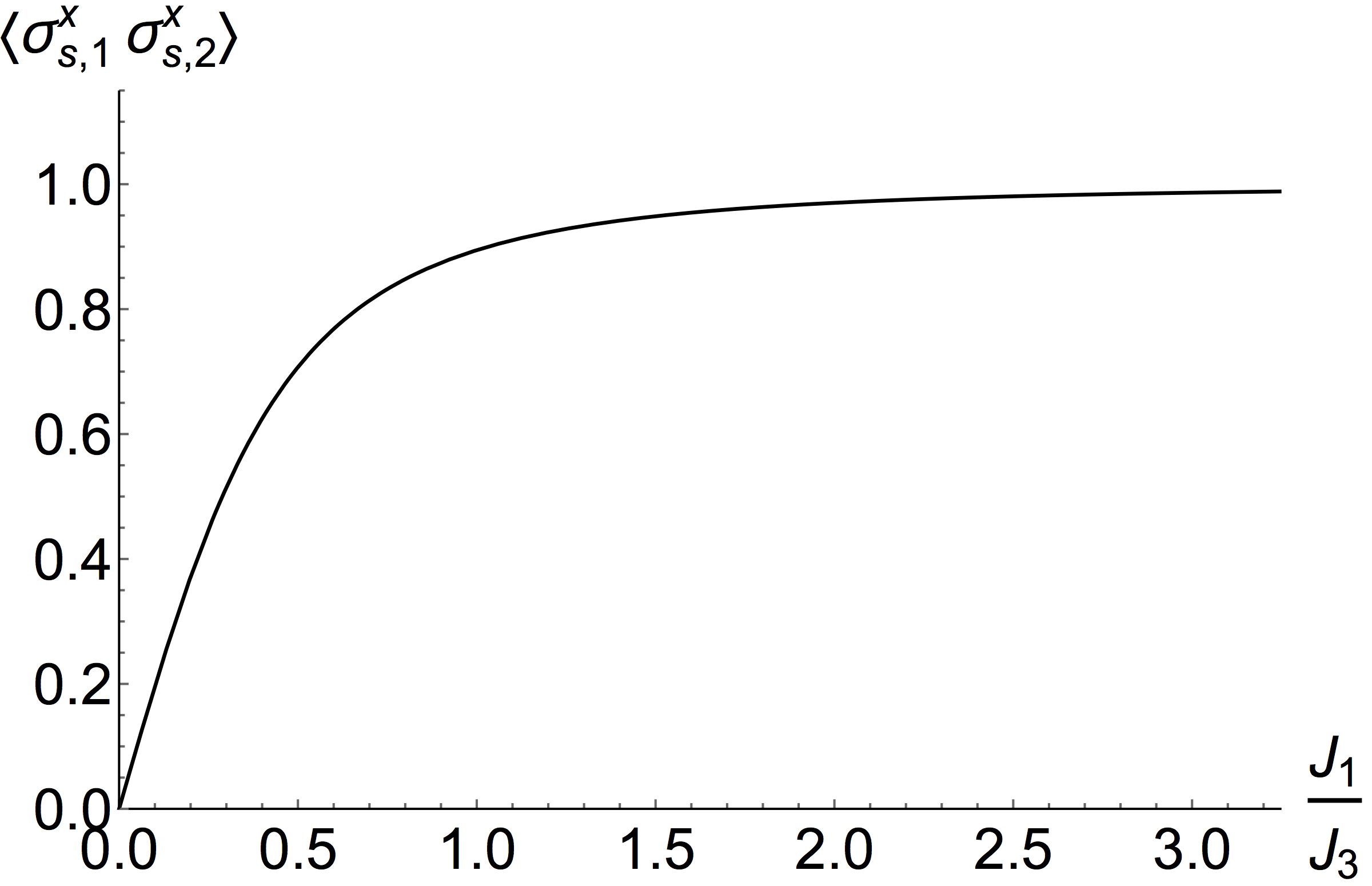} 
			\caption{Spin polarization between two nearest sites when $J_2 = J_4 = 0$.}
	\label{fig:sigma_x}
\end{figure}

Taking into account that the original Hamiltonian is doubled in our case, we get
\begin{eqnarray}
\mathcal{N}^2 &=& \frac{1}{2} \sum_{i=1}^{4} \left| x_i\right|^2 \\ \nonumber
&=& \frac{1}{2}\left( e^2 + \left| \alpha \right|^2 + \frac{1}{\left|\beta \right|^2 e^2} \left( e^4 - \left|\alpha \right|^4\right) \left( e^2 - \left|\alpha \right|^2\right) \right).
\end{eqnarray}
We select two sites in $s$-th unit cell (see Fig.~\ref{fig:ribbon}) and calculate the spin correlation in the $x$ direction:
\begin{gather*}
\left< \sigma_{s,1}^x \sigma_{s,2}^x\right> = -i \left< c_{s,1}c_{s,2}\right> = \frac{-i}{M} \sum_{q,q'} e^{iqr_{s,1}+iq'r_{s,2}} \left< c_{q,1}c_{q',2}\right>.
\end{gather*}
We find:
\begin{eqnarray}
\left\{ c_{k,\lambda}, c_{k', \lambda'} \right\} &=& 2\delta_{k,-k'}\delta_{\lambda,\lambda'}, \\ \nonumber
 \left< c_{q,1}c_{q',2}\right> &= & \frac{1}{\mathcal{N}^2} \sum_{k,k'} \left( -x_1x_2^* \delta_{q',-k'}\delta_{q,k} + x_1^*x_2 \delta_{q,-k'}\delta_{q',k}\right).
  \end{eqnarray}
 In the $r_{s,1} \to \infty$ limit,
\begin{eqnarray}
\left< \sigma_{s,1}^x \sigma_{s,2}^x\right> &=& \frac{2e\left| J_1 \right|}{M\mathcal{N}^2} \sum_{k,k'} e^{(k-k')\cdot r_{s,1}} 
\\ \nonumber
&=& \frac{2e\left| J_1 \right|}{M\mathcal{N}^2} \sum_{k,k'} \delta_{k,k'} = \frac{2e\left| J_1\right|}{\mathcal{N}^2}.
\end{eqnarray}
Then, we get an analytical expression for the spin correlation:
\begin{eqnarray}
\left< \sigma_{s,1}^x \sigma_{s,2}^x\right> = 4\left( \frac{g}{t} + \frac{t}{g} + \left( \frac{g}{t} - \left( \frac{t}{g}\right)^3\right)\left(g^2-t^2\right)\right)^{-1}.
\end{eqnarray}
From Fig.~\ref{fig:sigma_x}, we check that the spin polarization along the $x$ direction changes continuously from $0$ to $1$ when we increase the value of $\left|J_1\right|$. 

\section{Renormalization Group Analysis}

In this Appendix, for completeness, we present the results of the Renormalization Group (RG) analysis of the $b_{12}$ Sine-Gordon term \cite{Giamarchi} in Eq. (30). We will use the following result, which holds for primary (charge) fields of Gaussian models described by a Luttinger theory
(and the Hamiltonian $H_g$ in Eq. (32))  
\begin{equation}
\langle\phi(x)\phi(0)\rangle-\langle\phi^2(0)\rangle=\frac{K}{4}\ln\left(\frac{l^2}{l^2+x^2}\right),
\end{equation}
where $l$ is the lattice spacing, and can be understood as a short-distance cutoff parameter. We have rescaled $\phi$ as $\sqrt{K}\phi$, to make an analogy with free fermion models.

We will use this property for the $\phi_-(x)$ field defined in the main text in Sec. IVB.  The idea is to study the RG flow of the parameter $b_{12}$ by changing the lattice spacing $l$ and imposing that the participation function of the system remains unchanged.  The partition function is 
\begin{equation}
{\cal Z}=Tr(e^{-\beta H})=Tr(e^{-\beta (H_g+H_{b_{12}})}).
\end{equation}
We suppose $H_{b_{12}}\ll H_g$ (which is true at short distances or in the ultraviolet limit for energies $\sim |J_1|$). Thus, we can factorize the partition function : ${\cal Z}={\cal Z}_0\langle Tr(e^{-\beta H_{b_{12}}})\rangle_0$, 
with ${\cal Z}_0=Tr(e^{-\beta H_g})$ the partition function of the massless free field. 

Now, we expand the exponential in the partition function. The first order term is equal to zero, since at high energy the phase $\phi_-(x)$ fluctuates randomly in $[0,2\pi]$ and thus $\langle \cos(\phi_-)\rangle_0=0$. To second order, we find 
\begin{eqnarray}
{\cal Z}^{(2)} &=& {\cal Z}_0\int_0^{\beta} d\tau\int_0^{\beta} d\tau'\int dx\int dx' \times \\ \nonumber
&&\frac{b_{12}(l)^2}{l^4}\langle\cos(\sqrt{8}\phi_-(x,\tau))\cos(\sqrt{8}\phi_-(x',\tau'))\rangle_0,
\end{eqnarray}
with 
\begin{eqnarray}
&&\langle\cos(\sqrt{8}\phi_-(x,\tau))\cos(\sqrt{8}\phi_-(x',\tau'))\rangle_0 \\ \nonumber
&\approx& \left(\frac{l^2}{|x-x'|^2+v^2|\tau-\tau'|^2}\right)^{2K_-},
\end{eqnarray}
and $l\ll (|x-x'|,v|\tau-\tau'|)$ by hypothesis.
Now, we increase the scaling parameter $l\rightarrow l'=le^{\frac{dl}{l}}=le^{d\lambda}$, with $d\lambda=\frac{dl}{l}$, and impose that the partition function should stay constant : ${\cal Z}(l)={\cal Z}(l')$. This condition yields :
\begin{eqnarray}
\ln(b_{12}(l')-\ln(b_{12})(l)=(2-2K_-)\ln\left(\frac{l'}{l}\right).
\end{eqnarray}
We observe that for $K_-<1$, the dimensionless parameter $b_{12}$ will grow under RG. The Sine-Gordon term will become as important as the Gaussian theory characterized by the velocity $v=|J_1l|\sim 1$,
roughly when $b_{12}(l_c)\sim 1$, which corresponds to the limit of validity of the Gaussian model. This implies $l_c=\frac{1}{|J_1|}\left(\frac{J_1}{J_3}\right)^{\frac{1}{2-2K_-}}$.
This allows us to define the mass (or gap) associated with the mode $\phi_-$:
\begin{equation}
m^*\sim \frac{1}{l_c}\sim |J_1|\left(\frac{J_3}{J_1}\right)^{\frac{1}{2-2K_-}}.
\end{equation}
In the low energy regime, $l_c\ll l$, the massive mode $\phi_-$ is locked in the ground state, in order to minimize the energy.
Note that, as soon as $J_1\neq J_2$, a gap $\Delta\sim |J_1-J_2|$ opens and dominates in front of $m^*$.

We can now deduce the spin-spin correlation functions $\langle\sigma_{\alpha}^z(x)\sigma_{\alpha}^z(0)\rangle=\langle a^{\dagger}_{\alpha}(x)a_{\alpha}(x)a^{\dagger}_{\alpha}(0)a_{\alpha}(0)\rangle$.
For example, let us consider the chain $\alpha=1$. Then:
\begin{equation}
a_1^{\dagger}(x)a_1(x)=\frac{1}{\sqrt{2}}\partial_x(\phi_++\phi_-)+e^{i2xk_F}e^{i\sqrt{2}(\phi_++\phi_-)}.
\end{equation}
Since $\phi_-$ is locked, we have $\partial_x \phi_-(x)=0$, and then:
\begin{eqnarray}
\langle \partial_x\phi_+(x)\partial_x\phi_+(0)\rangle 
&=&-K_+\partial_{xx}^2\left[\frac{1}{4}\ln\left(\frac{l^2}{l^2+x^2}\right)\right] \nonumber \\
&\approx& \frac{K_+}{x^2}.
\end{eqnarray}
In a similar way, we find:
\begin{eqnarray}
\langle e^{i2xk_F}e^{i\sqrt{2}\phi_+(x)} e^{-i\sqrt{2}\phi_+(0)}\rangle  \nonumber \\
&=& \left(\frac{l^2}{l^2+x^2}\right)^{K_+/2}(-1)^{x/l}  \nonumber \\
&\sim& \frac{(-1)^{x/l}}{x^{K_+}}.
\end{eqnarray}
Finally, we find Eq. (35):
\begin{equation}
\langle\sigma_1^{z}(x)\sigma_1^z(0)\rangle \sim \frac{K_+}{x^2}+\frac{(-1)^{x/l}}{x^{K_+}}.
\end{equation}
 For the single chain, the $z$ component of spin correlation functions decay as $1/x^2$.

\end{document}